\documentclass[aps,prb,reprint,superscriptaddress]{revtex4-1}

\usepackage[pdftex]{graphicx}
\usepackage{color}
\usepackage{enumitem}
\usepackage{gensymb}
\usepackage{epstopdf}
\usepackage{amsmath}

\begin{document}


\title{Magnetic relaxation phenomena in the chiral magnet Fe$_{1-x}$Co$_x$Si: An ac susceptibility study}



\author{L.J. Bannenberg}
\affiliation{Faculty of Applied Sciences, Delft University of Technology, Mekelweg 15, 2629 JB Delft, The Netherlands}
\author{A.J.E. Lefering}
\affiliation{Faculty of Applied Sciences, Delft University of Technology, Mekelweg 15, 2629 JB Delft, The Netherlands}
\author{K. Kakurai}
\affiliation{Neutron Science and Technology Center, CROSS Tokai, Ibaraki 319-1106, Japan}
\affiliation{RIKEN Center for Emergent Matter Science(CEMS), Wako 351-0198, Japan}
\author{Y. Onose}
\affiliation{Department of Basic Science, University of Tokyo, Tokyo, 153-8902, Japan}
\author{Y. Endoh}
\affiliation{RIKEN Center for Emergent Matter Science(CEMS), Wako 351-0198, Japan}
\author{Y. Tokura}
\affiliation{RIKEN Center for Emergent Matter Science(CEMS), Wako 351-0198, Japan}
\affiliation{Department of Applied Physics, University of Tokyo, Tokyo 113-8656, Japan}
\author{C. Pappas}
\affiliation{Faculty of Applied Sciences, Delft University of Technology, Mekelweg 15, 2629 JB Delft, The Netherlands}

\date{\today}

\begin{abstract}
We present a systematic study of the ac susceptibility of the chiral magnet Fe$_{1-x}$Co$_x$Si with $x$ = 0.30 covering four orders of magnitude in frequencies from 0.1 Hz to 1 kHz, with particular emphasis to the pronounced history dependence. Characteristic relaxation times ranging from a few milliseconds to tens of seconds are observed around the skyrmion lattice A-phase, the helical-to-conical transition and in a region above $T_C$. The distribution of relaxation frequencies around the A-phase is broad, asymmetric and originates from multiple coexisting relaxation processes. The pronounced dependence of the magnetic phase diagram on the magnetic history and cooling rates as well as the asymmetric frequency dependence and slow dynamics suggest more complicated physical phenomena in Fe$_{0.7}$Co$_{0.3}$Si than in other chiral magnets.

\end{abstract}
\pacs{}

\maketitle
\section{\label{sec:level1} Introduction}
The discovery of skyrmion lattices in cubic helimagnets such as MnSi,\cite{muhlbauer2009} FeGe,\cite{FeGe_Lorenz_TEM,moskvin2013} Cu$_2$OSeO$_3$\cite{seki2012observation,seki2012formation,adams2012long} and Fe$_{1-x}$Co$_x$Si\cite{yu2010,munzer2010} has increased the interest in non-centrosymmetric magnetic materials with Dzyaloshinsky-Moriya interactions.\cite{D, M} The skyrmion lattice phase is a periodic array of spin vortices observed in the A-phase, a region in the magnetic field ($B$) - temperature ($T$) phase diagram below the critical temperature $T_C$, and was first observed in reciprocal space in a single crystal of MnSi by neutron scattering \cite{muhlbauer2009} and subsequently in real space in a thin film of Fe$_{0.5}$Co$_{0.5}$Si by Lorentz Transmission Microscopy.\citep{yu2010} 

The chiral skyrmions that form the skyrmion lattice are non-coplanar and topologically stable spin textures with dimensions significantly larger than the inter-atomic distances.\cite{bogdanov1989,bogdanov1994,Rossler2006,nagaosa2013} The potential application of skyrmions as low-current high-density information carriers and in other spintronic devices as well as the unexplored magnetic properties drive the scientific interest in this non-conventional magnetic ordering.\cite{nagaosa2013,fert2013,romming2013,oike2016}

The helical order at zero field is stabilized in the pseudo-binary compound Fe$_{1-x}$Co$_x$Si in a wide range of chemical substitution of 0.05 $<$ $x$ $<$ 0.8.\cite{beille1981,beille1983,motokawa1987} As for other members of the B20 group such as MnSi and Cu$_2$OSeO$_3$, this helical order is a result of the balance between the strong ferromagnetic and the weaker Dzyaloshinsky-Moriya (DM) interactions that arises from the non-centrosymmetric crystal structure. The $B$ - $T$ phase diagrams of Fe$_{1-x}$Co$_x$Si compounds are quantitatively similar to each other and to the other B20 group members. Below $T_C$, three ordered states dominate the phase diagram: a helical phase occurring at low fields $B$ \textless $B_{C1}$, where the weak anisotropy fixes the orientation of the helices typically along the $\langle$100$\rangle$ or $\langle$111$\rangle$ crystallographic directions, a conical phase for intermediate fields $B_{C1}$ $<$ $B$ $<$ $B_{C2}$, where the magnetic field overcomes the anisotropy and orients the helices along the magnetic field,  and the A-phase close to $T_C$, where the skyrmion lattice phase is stabilized. Magnetic fields exceeding $B_{C2}$ overcome the DM interactions inducing a field-polarized state.  

The ability to tune important physical properties by chemical substitution as well as the high degree of chemical disorder make Fe$_{1-x}$Co$_x$Si of particular interest among the B20 compounds.\cite{onose2005} In particular, the amount of chemical substitution changes both the sign and the magnitude of the DM-interactions. It thus affects the $T_C$ that ranges from a few Kelvin to 50 K and the magnetic chirality which changes from left handed to right-handed at $x$ = 0.65.\citep{siegfried2015} Additionally, it alters the pitch of the helical ordering from $\sim$30~nm to $\sim$200~nm as this pitch is proportional to the ratio of the ferromagnetic exchange to the DM interactions.\citep{beille1983,grigoriev2009}

Different from the archetype chiral system MnSi and other systems of the same family, Fe$_{1-x}$Co$_x$Si appears to have a phase diagram depending on the magnetic history\cite{munzer2010,bauer2016,bannenberg2016} and also on the applied cooling rates through $T_C$.\cite{bannenberg2016} Additionally, neutron scattering shows that skyrmion lattice correlations may persist down to the lowest temperatures depending on the magnetic history of the sample.\cite{munzer2010,milde2013,bannenberg2016} Several DC magnetization\cite{beille1981,watanabe1985,motokawa1987,chattopadhyay2002,grigoriev2007,bauer2016} and some AC susceptibility studies have been performed so far, but only with an AC drive frequency of 30 Hz\cite{ou-yang2015} and 1000 Hz.\cite{onose2005,bauer2016} Based on these studies, phase-diagrams have been deduced for a wide range of chemical substitution and field directions but no attention has been devoted to the frequency dependence.\cite{ou-yang2015,bauer2016} 

The AC susceptibility measurements presented here for Fe$_{0.7}$Co$_{0.3}$Si complement previous neutron scattering as well as AC susceptibility studies as they span a broad frequency range of four orders in magnitude, from 0.1~Hz to 1000~Hz and have a particular emphasis on the influence of the magnetic hysteresis and the applied cooling rate. The results confirm the history dependence reported earlier and show a strong dependence of the imaginary component of the AC susceptibility on the AC drive frequency around the A-phase, helical-to-conical transition and in a region above $T_C$. Around the A-phase, the distribution of relaxation frequencies is found to be broad and asymmetric, indicating the occurrence of multiple coexisting very slow relaxation processes.

The remainder of this paper is organized as follows. Section II discusses the experimental details, Section III the Zero Field Cooled (ZFC) AC susceptibility study at a frequency of 5 Hz and Section IV the magnetic history and cooling rate dependence and the $B$ - $T$ phase diagrams for both ZFC and Field Cooling (FC) at a frequency of 5 Hz. Section V confers the frequency dependence, Section VI shows $B$ - $T$  and phase diagrams for $f$ = 0.1, 5 and 100~Hz after ZFC. Conclusions are given in Section VII.

\section{\label{sec:level2} Experimental Details}
The measurements were performed on a 20 mg single crystal of Fe$_{0.7}$Co$_{0.3}$Si originating from the same batch as the crystal studied previously by neutron scattering.\cite{takeda2009,bannenberg2016}. The crystal quality was tested with Laue X-ray diffraction and it was aligned with the [110] direction vertical within $\pm$ 10\degree. The sample has an irregular shape and its longest direction was roughly vertically oriented. 

The real $\chi^\prime$ and imaginary $\chi^\prime$$^\prime$ components of the AC susceptibility were measured with a MPMS-XL Quantum Design SQUID magnetometer using the extraction method. The DC field was applied along the vertical axis and the AC field of 0.1 $\leq$ $B_{AC}$ $\leq$ 0.4~mT was oriented parallel to the DC field. Several measurements in and around the A-phase at $T$ = 41 K revealed that the susceptibility was independent of the ac field and subsequent measurements where performed with $B_{ac}$ = 0.4~mT. Three specific protocols that are similar to those adopted for the previous neutron scattering experiment\cite{bannenberg2016} have been used:
\begin{itemize}
\item ZFC temperature scans: the sample was cooled from 60~K to 6~K under zero field. Then a magnetic field was applied and the signal was recorded by stepwise increasing the temperature. The system was brought to thermal equilibrium before the measurement at each temperature commenced. 
\item FC temperature scans: the sample was brought to 60~K where a magnetic field was applied. The temperature was then decreased stepwise and the measurements commenced once the system reached the thermal equilibrium. 
\item Fast FC temperature scans: the sample was brought to 60~K where a magnetic field was applied. The temperature was then decreased with 10 Kmin$^{-1}$ to 30~K. Subsequently, the signal was recorded by decreasing the temperature stepwise and after waiting for the system to reach thermal equilibrium. 
\end{itemize}

\section{\label{sec:level3} ZFC ac susceptibility at 5 Hz}
\begin{figure}
\begin{center}
\includegraphics[width= .45 \textwidth]{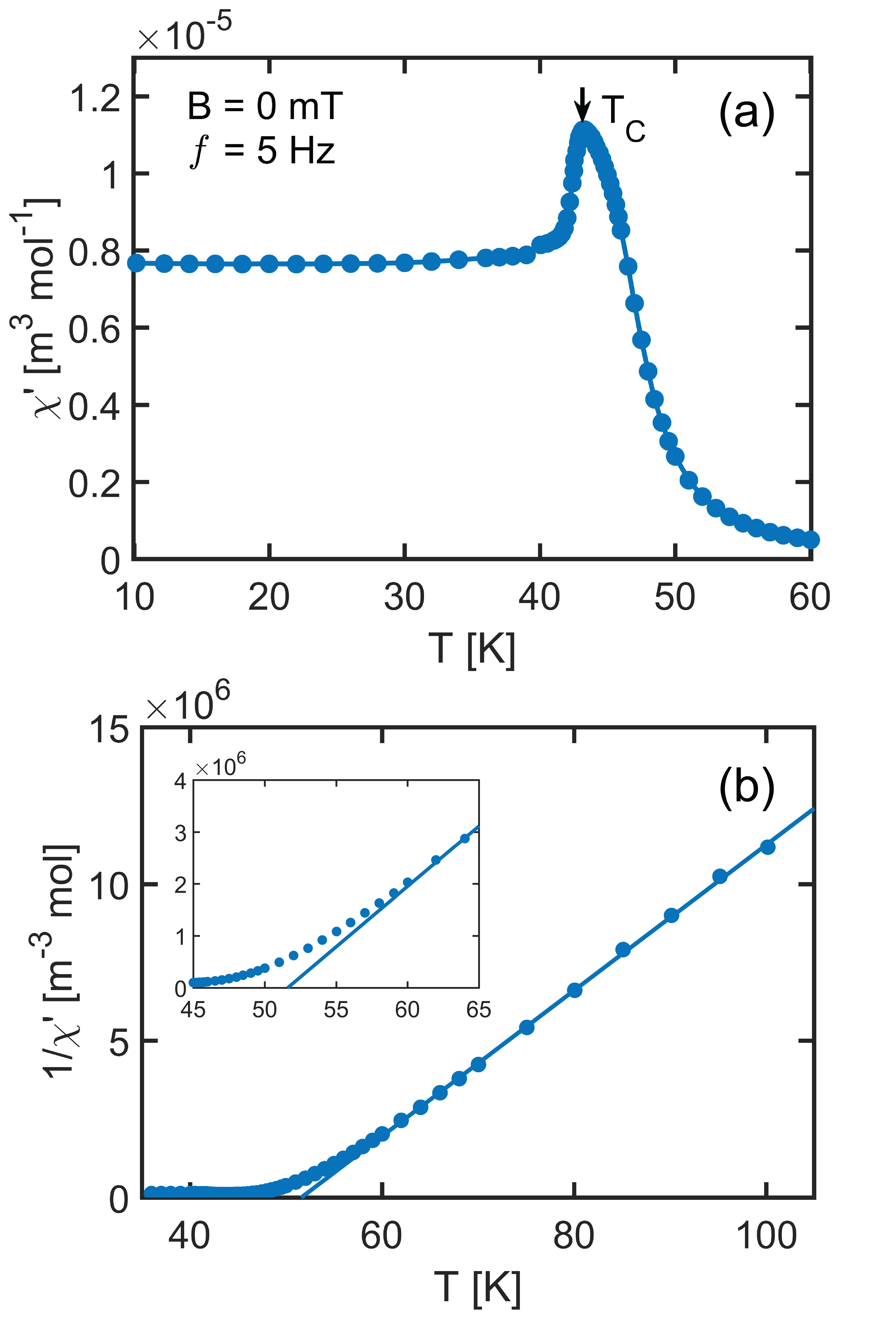}
\caption{Zero field susceptibility for Fe$_{0.7}$Co$_{0.3}$Si measured at $f$ = 5~Hz. (a) Temperature dependence of $\chi^\prime$. (b) Inverse susceptibility that has been fitted to a Curie-Weiss law.}
\label{ZeroField}
\end{center}
\end{figure}

Fig. \ref{ZeroField}(a) depicts the temperature dependence of $\chi^\prime$ and $\chi^\prime$$^\prime$ at $B$ = 0~mT and $f$ = 5~Hz. A maximum in $\chi^\prime$ at $T_C$ $\approx$ 43.2~K marks the transition to the helical order, which is characterized by a pitch of $\ell$ $\sim$~40~nm.\citep{takeda2009,bannenberg2016} In the helical ordered phase, $\chi^\prime$ drops by about 30\% from its maximum value and remains almost constant for temperatures below 40 K, which is quantitatively similar to the behavior reported in the literature.\cite{onose2005} 

At higher temperatures, the susceptibility follows a Curie-Weiss behavior as can be inferred from the linear relation of the inverse susceptibility with temperature displayed in Fig. \ref{ZeroField}(b). The corresponding fit with the Curie-Weiss law  $\chi^{\prime} = C/(T-T_{CW})$ reveals a Curie-Weiss temperature of $T_{CW}$ = 51.6 $\pm$ 0.5~K and a Curie constant of 4.3 $\pm$ 0.1 m$^3$ mol$^{-1}$ $\times$ 10$^{-5}$, which translates to 1.4 $\mu_B$ f.u.$^{-1}$ and is as such in good agreement with the literature.\citep{bauer2016} Deviations from the Curie-Weiss law occur for temperatures below 62~K, i.e. approximately 1.4 $T_C$.

\begin{figure*}
\begin{center}
\includegraphics[width= 1 \textwidth]{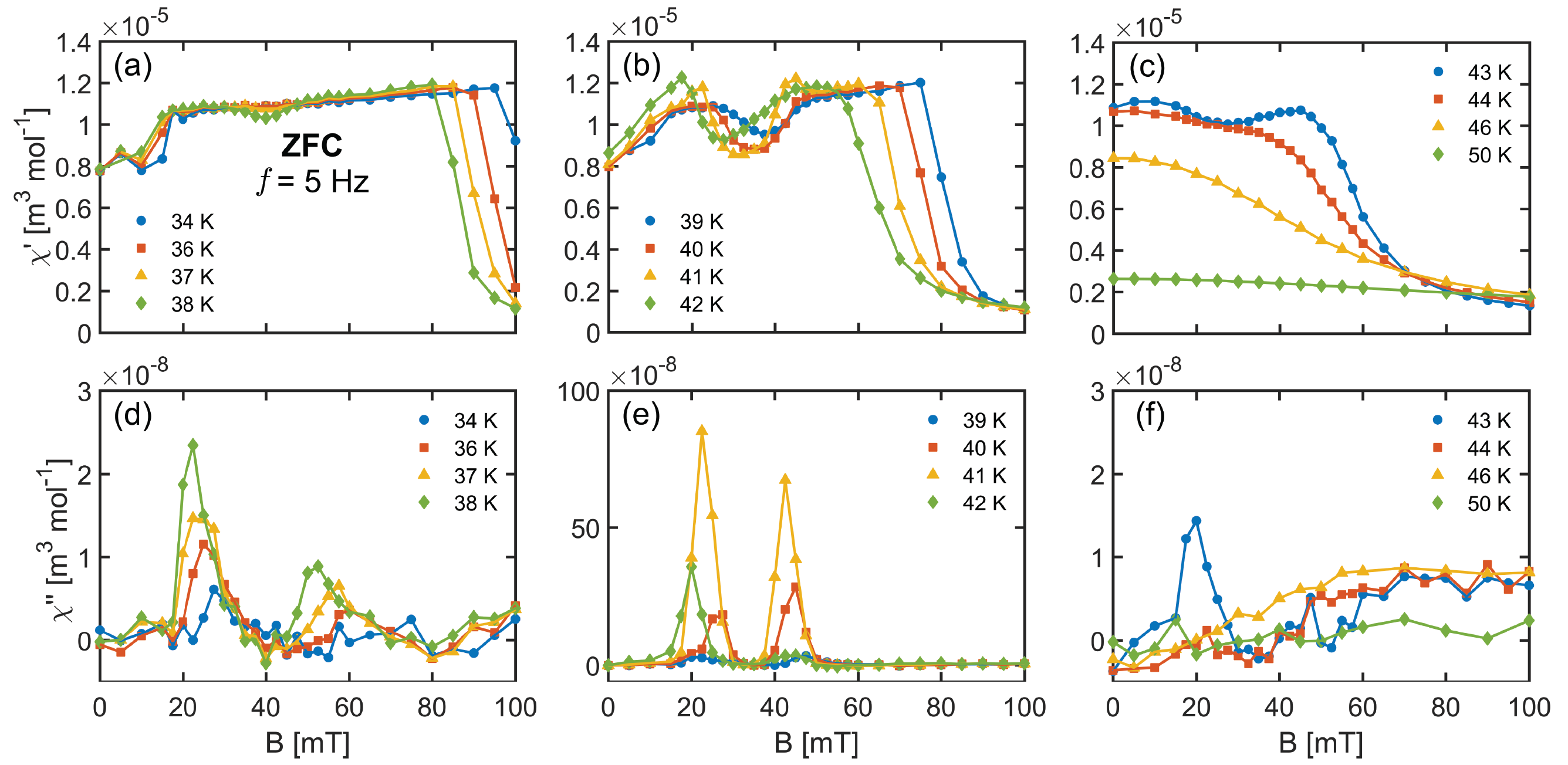}
\caption{Magnetic field dependence of (a) - (c) $\chi^\prime$ and (d) - (f) $\chi^\prime$$^\prime$ of Fe$_{0.7}$Co$_{0.3}$Si at $f$ = 5~Hz for the temperatures indicated. The field was applied after zero field cooling.}
\label{temp5Hz}
\end{center}
\end{figure*}
\begin{figure}
\begin{center}
\includegraphics[width= .5 \textwidth]{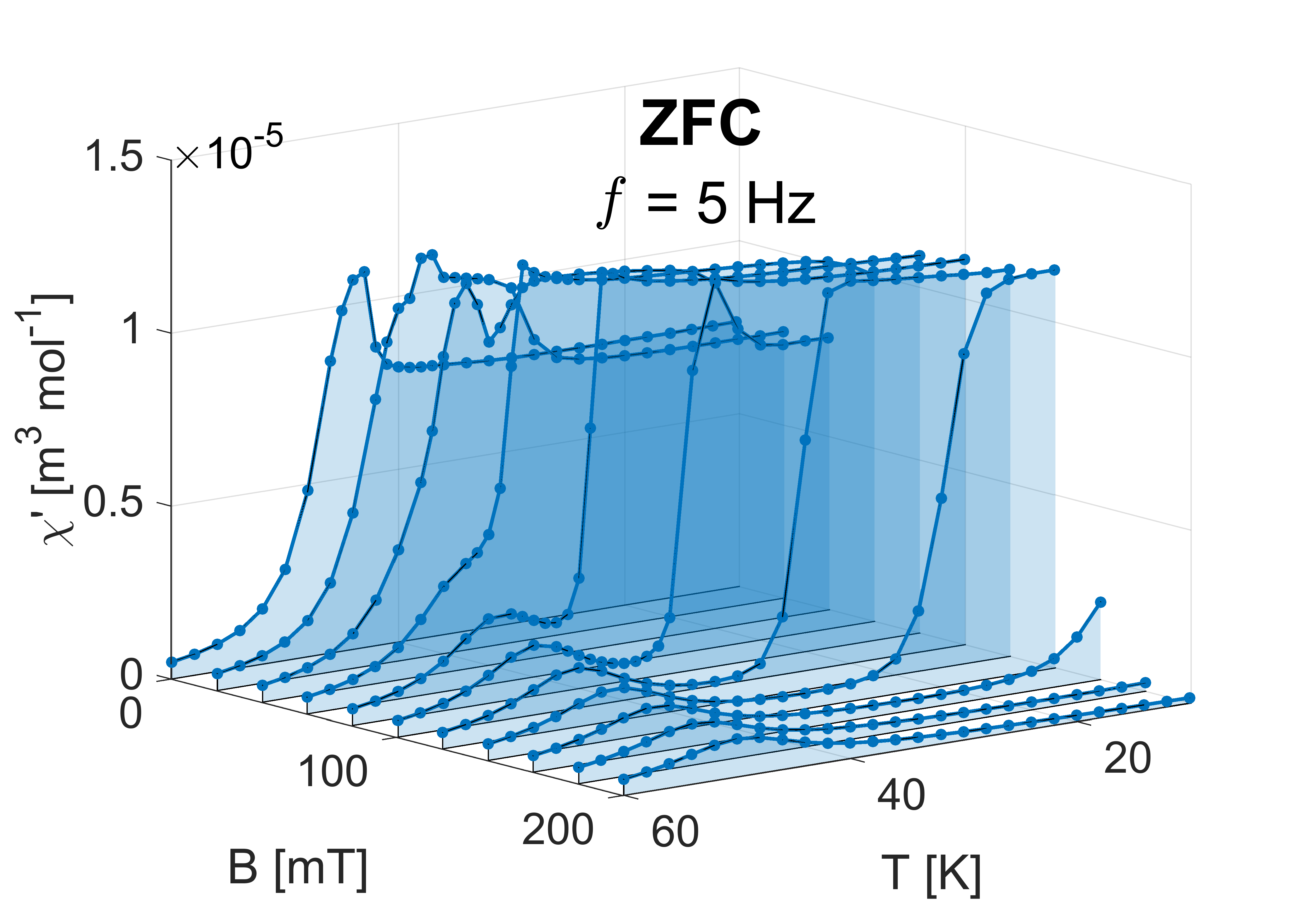}
\caption{Temperature dependence of $\chi^\prime$ of Fe$_{0.7}$Co$_{0.3}$Si for several magnetic fields at $f$ = 5~Hz. The field was applied after zero field cooling.}
\label{waterfall_5Hz}
\end{center}
\end{figure}

\begin{figure}
\begin{center}
\includegraphics[width= .48 \textwidth]{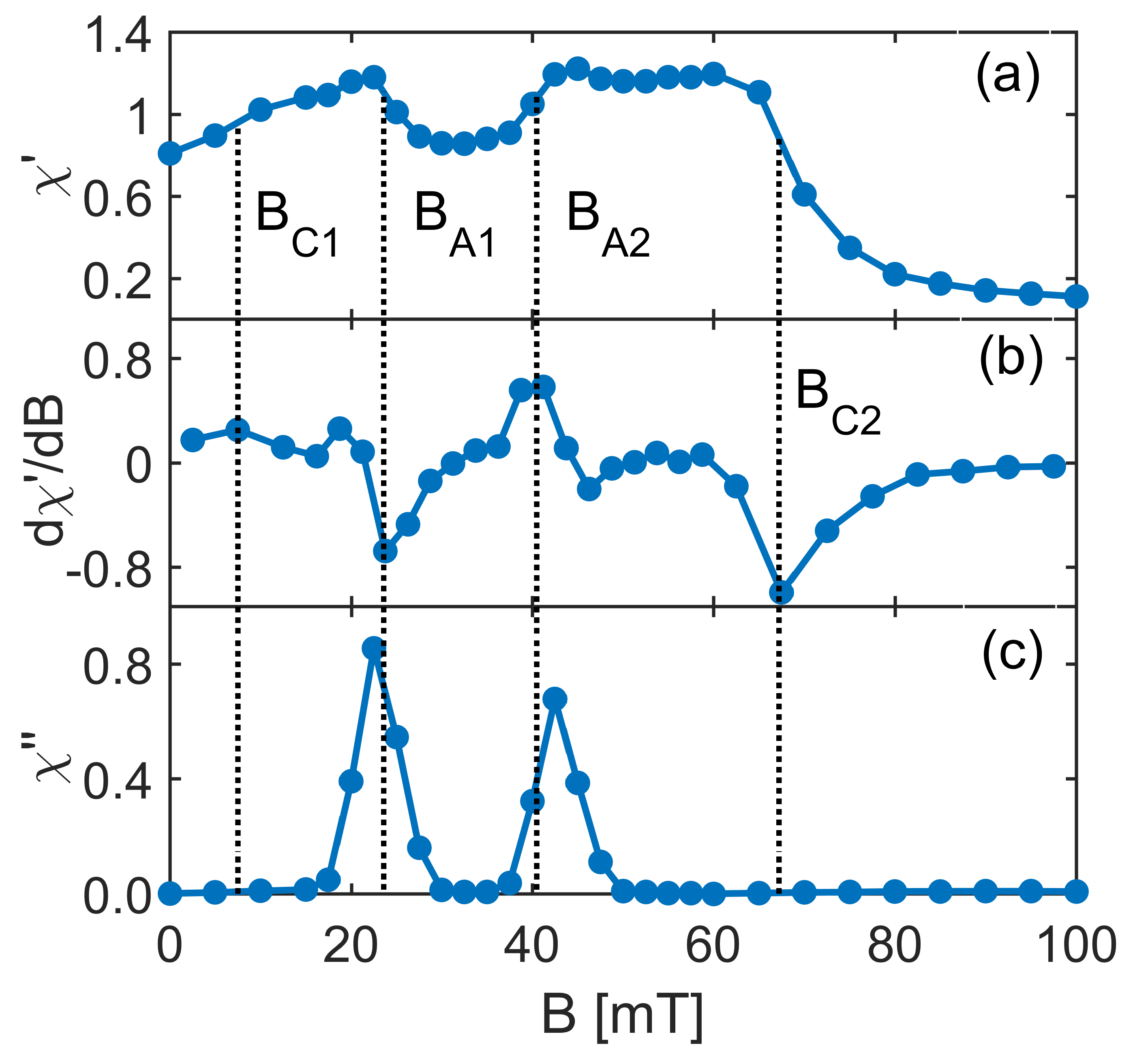}
\caption{Magnetic field dependence at $T$ = 41 K of (a) $\chi^\prime$ in units of m$^3$mol$^{-1}$$\times 10^{-5}$, (b) $d\chi^\prime/dB$ in units of m$^3$mol$^{-1}$T$^{-1}$$\times 10^{-3}$ and (c) $\chi^{\prime\prime}$ in units of m$^3$mol$^{-1}$$\times 10^{-6}$. The field was applied after ZFC. The local maxima/minima of $d\chi^\prime/dB$ define the lower critical field $B_{C1}$, the higher critical field $B_{C2}$, as well as the lower and upper boundaries of the A-phase $B_{A1}$ and $B_{A2}$.}
\label{Fig_Phase_Boundaries_41K}
\end{center}
\end{figure}

\begin{figure*}
\begin{center}
\includegraphics[width= 1 \textwidth]{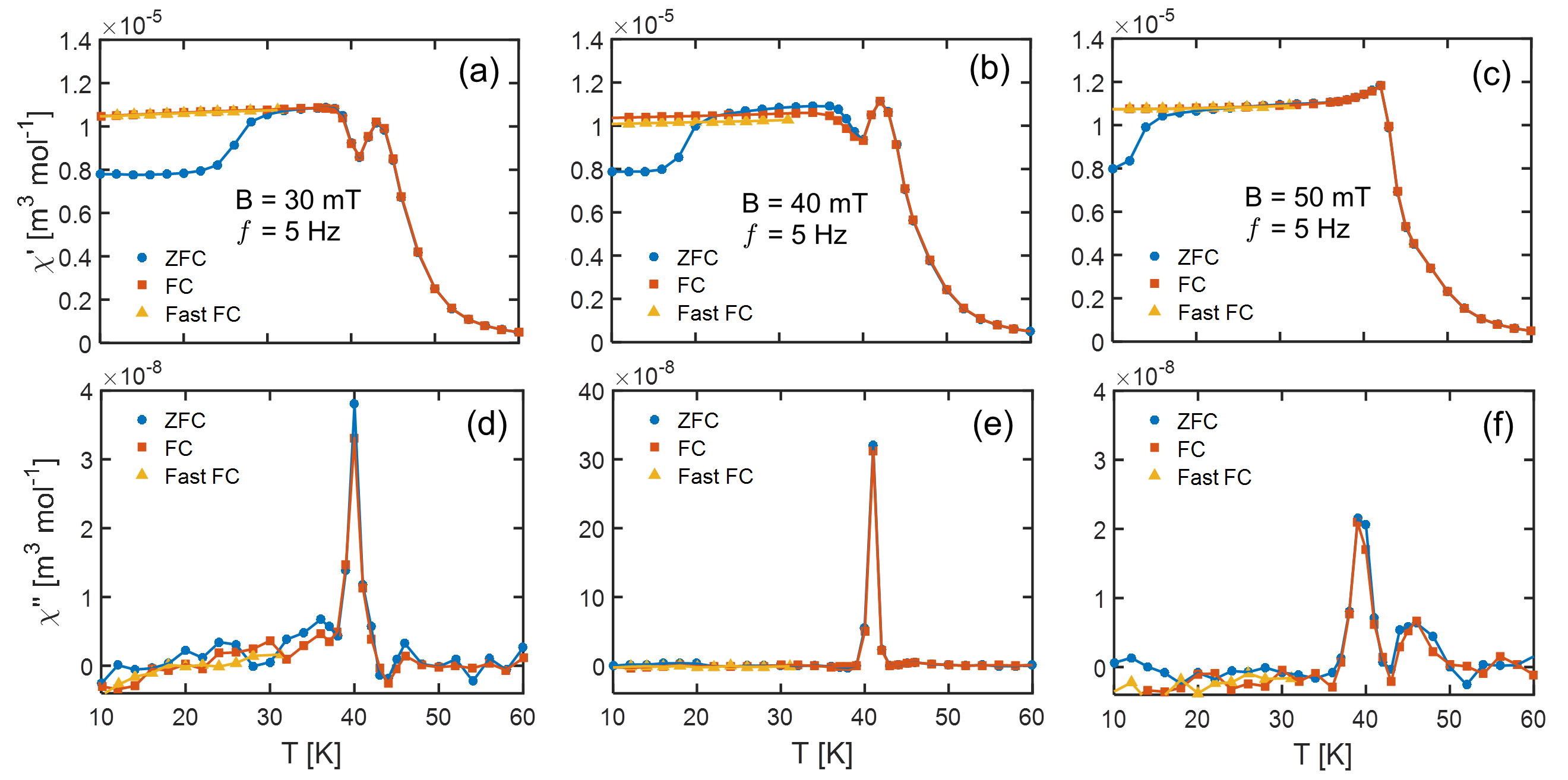}
\caption{Temperature dependence of (a) - (c) $\chi^\prime$ and (d) - (f) $\chi^\prime$$^\prime$ of Fe$_{0.7}$Co$_{0.3}$Si at $f$ = 5~Hz for the magnetic fields indicated. The magnetic field was applied after Zero Field Cooling (ZFC), Field Cooling (FC) and Fast Field Cooling (Fast FC).}
\label{hist}
\end{center}
\end{figure*}

To evaluate the magnetic field dependence of the susceptibility, we display ZFC $\chi^\prime$ in Figs. \ref{temp5Hz} (a)-(c) for various temperatures. For $T$ $<$ 38~K, Fig. \ref{temp5Hz} (a) shows that $\chi^\prime$ increases with increasing magnetic field until the relatively low field of $B_{C1}$ of the helical-to-conical transition. For 38~K $\leq$ $T$ $\leq$ 44~K, Fig. \ref{temp5Hz} (b) reveals a clear dip in $\chi^\prime$ between $\sim$20 - 45~mT which is maximal at 41~K. This dip marks the A-phase with its characteristic skyrmion lattice correlations. Fig. \ref{temp5Hz} (c) shows that above 44~K, $\chi$$^\prime$ decreases monotonically with increasing magnetic field as well as with increasing temperature for a given magnetic field. 

The corresponding evolution of $\chi^\prime$$^\prime$ is displayed in Figs. \ref{temp5Hz} (d)-(f) as a function of magnetic field. In these figures, two peaks around $\sim$20 and $\sim$45 mT are visible of which the latter one is slightly less intense. These peaks mark the boundary of the A-phase and appear for 34~K $\leq$ $T$ $\leq$ 43~K and thus over a much broader temperature range than the dip in $\chi^\prime$, which is only seen between $T$ = 38~K and $T$ = 42~K. Similar peaks in $\chi^\prime$$^\prime$ have also been observed at the boundary of the A-phase for MnSi\cite{bauer2012}and Cu$_2$OSeO$_3$,\cite{qian2016} but in a considerably less wide temperature range.

Another, although small, increase in $\chi^\prime$$^\prime$ is observed at high fields and is observed clearly in Fig. \ref{temp5Hz} (f). Additional measurements show that this feature does not disappear even at fields as high as 1~T.  This non-zero $\chi^\prime$$^\prime$ persists above $T_C$ up to 50~K and for fields exceeding 40~mT and has a broad maximum around 46 K. A similar effect has also been reported for Cu$_2$OSeO$_3$\cite{qian2016} and a more elaborate discussion of this feature, including its frequency dependence, will be presented in Section VI.  

An overview of the temperature and magnetic field dependence of $\chi^\prime$ is given in Fig. \ref{waterfall_5Hz}. A well-defined maximum is visible at $T_C$ at zero field. At 40 mT, the dip characteristic for the A-phase is visible slightly below $T_C$. At a field $B$ $\sim$ 60~mT, a kink appears at $T$ = 46~K, which marks the split of the single maximum for fields $B$ $<$ 60~mT into two separate maxima for $B$ $>$ 60~mT. The low temperature maximum is related to the DM interaction and shifts to lower temperatures for increasing magnetic fields and marks the $B_{C2}$ transition from the conical to the field polarized state. The high temperature maximum reflects the ferromagnetic correlations and shifts to higher temperatures for increasing fields. 

A similar behavior has been reported for other chiral magnets as MnSi,\cite{thessieu1997} FeGe,\cite{wilhelm2011} and Cu$_2$OSeO$_3$.\cite{zivkovic2014} In these studies, the high temperature maximum has been interpreted as a smeared transition from the high-temperature paramagnetic to the low-temperature field polarized state\cite{thessieu1997,wilhelm2011} and reflecting the classical ferromagnetic transition that would take place in the absence of DM interactions.\cite{zivkovic2014} 

\begin{figure*}
\begin{center}
\includegraphics[width= .8 \textwidth]{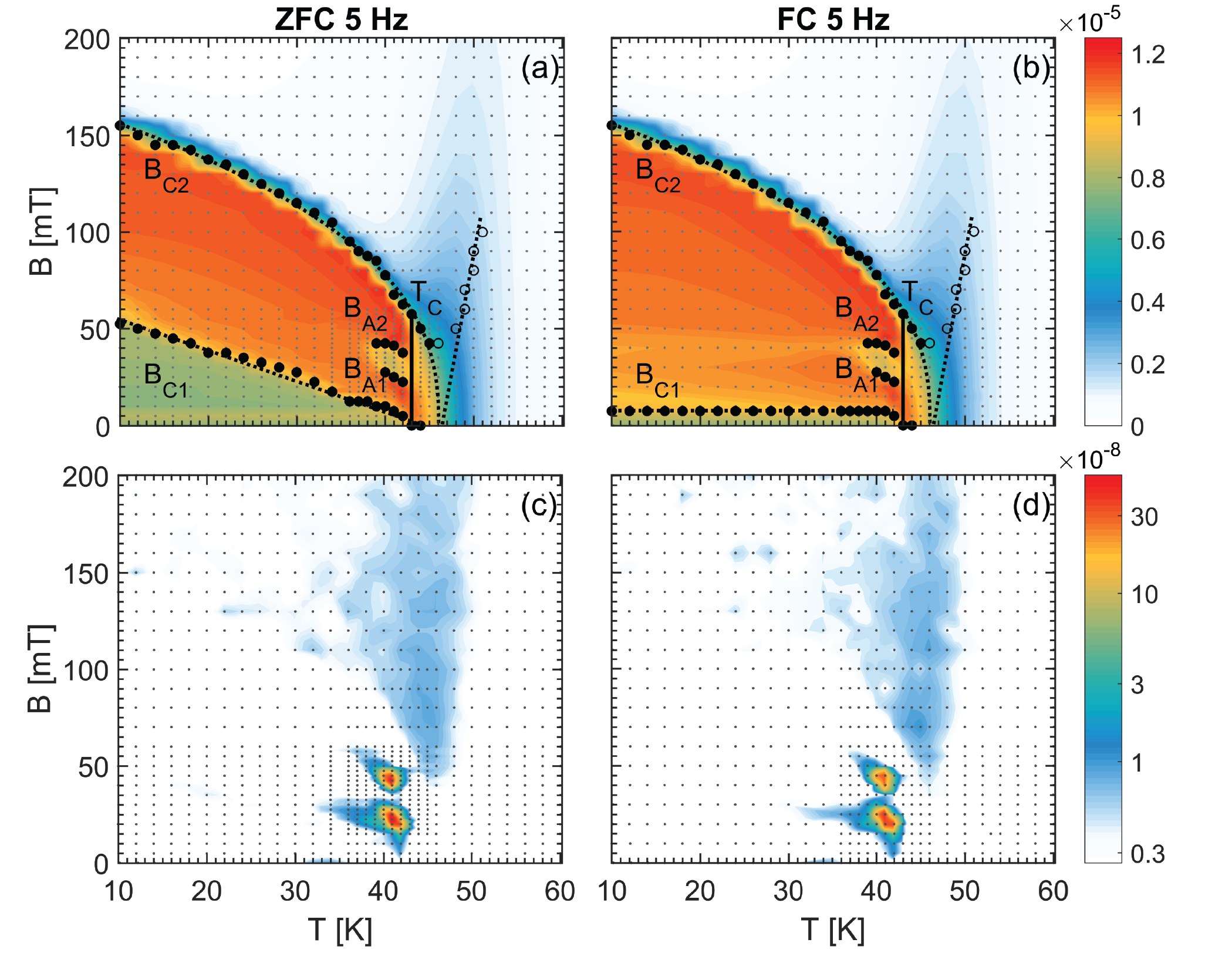}
\caption{Contour plots showing (a) - (b) $\chi^\prime$ and (c) - (d) $\chi^\prime$$^\prime$ in units of m$^3$mol$^{-1}$ after ZFC and FC at $f$ = 5 Hz as a function of temperature and magnetic field. $B_{C1}$, $B_{A1}$, $B_{A2}$ and $B_{C2}$ are defined by the inflection point of $\chi^\prime$ and are indicated with black circles. The dashed line along $B_{C2}$ indicate the fitted power law as described in the text. The grey dots indicate the points at which a signal has been recorded.}
\label{ZFC_FC}
\end{center}
\end{figure*}

To actually extract a magnetic phase diagram from the previously presented susceptibility results requires stringent criteria for the determinations of the critical fields $B_{C1}$ and $B_{C2}$ and the boundaries of the A-phase $B_{A1}$ and $B_{A2}$. Several criteria have been used in the literature, such as the maxima of $\chi^{\prime\prime}$ and the inflection points of $\chi^\prime$. As it has been discussed previously for MnSi\cite{bauer2012} and Cu$_2$OSeO$_3$,\cite{qian2016} different criteria lead to slightly different phase boundaries but not to significantly different physics. 

Fig. \ref{Fig_Phase_Boundaries_41K} displays the magnetic field dependence after ZFC of $\chi^\prime$, its derivative $d\chi^\prime/dB$  and $\chi^{\prime\prime}$ at $T$ = 41~K, a temperature where all the phase boundaries are present. At the borders of the A-phase, the extrema of $d\chi^\prime/dB$ which correspond to the inflection points of $\chi^\prime$, do not occur at exactly the same magnetic fields as the maxima of $\chi^{\prime\prime}$. In fact, the maxima of $\chi^{\prime\prime}$ lead to a slightly smaller value for $B_{A1}$ and a slightly higher one for $B_{A2}$ and thus to a larger pocket for the A-phase than the inflection points of $\chi^\prime$. A similar behavior has been found for both MnSi and Cu$_2$OSeO$_3$.\cite{bauer2012,qian2016} As $\chi^{\prime\prime}$ is almost zero at both $B_{C1}$ and $B_{C2}$, we choose the inflection points of $\chi^\prime$ to determine all the phase boundaries. 

\section{\label{sec:level4} History dependence at 5 Hz}
The previous section only discussed the susceptibility after ZFC. However, it is known that there is a strong hysteretic behavior of Fe$_{1-x}$Co$_{x}$Si\cite{munzer2010,bannenberg2016,bauer2016} that also depends on the cooling rate.\cite{bannenberg2016} For this reason, the ac susceptibility was also measured following the FC and Fast FC protocols described above. A selection of the results obtained with a frequency of 5 Hz is displayed in Figs. \ref{hist} (a)-(c) for $\chi^\prime$ and in Figs. \ref{hist} (d)-(f) for $\chi^{\prime^\prime}$ revealing a strong history dependence for $\chi^\prime$ but not for $\chi^{\prime\prime}$.

We start the discussion of the history dependence by comparing ZFC with FC. No differences exist above $T_C$. Below $T_C$, Figs. \ref{hist} (a) and (b) show a dip in $\chi^\prime$ centered around 40~K, which is due to the A-phase. This dip does not appear in Fig. 5 (c), under a magnetic field of 50~mT, which exceeds $B_{A2}$. 

At a lower temperature and for ZFC, $\chi^\prime$ drops from $\sim$ 1.1 m$^3$mol$^{-1}$$\times 10^{-5}$ to 0.8 m$^3$mol$^{-1}$$\times 10^{-5}$. This reflects the conical-to-helical transition which occurs at lower temperatures for higher magnetic fields. No substantial decrease in $\chi^\prime$ is observed for FC, implying that the conical phase extends to the lowest temperatures. 

Another difference between ZFC and FC $\chi^\prime$ is visible in Fig. \ref{hist} (b) at $B$ = 40~mT in the conical phase between $T$ = 20~K and 38~K. Here, $\chi^\prime$ is slightly lower after FC than after ZFC. This effect is enhanced for Fast FC. These differences are consistent with the previously reported neutron scattering results where it was shown that skyrmion lattice correlations persist in FC mode outside the A-phase and increase in intensity for higher cooling rates.\cite{bannenberg2016} Similarly to the A-phase, such  skyrmion lattice correlations would lead to a reduction of $\chi^\prime$, which is consistent with our observations.

The history, field and temperature dependence at 5~Hz of both $\chi^\prime$ and $\chi^\prime$$^\prime$ are summarized in the contour plots depicted in Fig. \ref{ZFC_FC}.  The helical phase shows up in the ZFC contour plot displayed in Fig. \ref{ZFC_FC} (a) below $B_{C1}$ and spans a wide section of the phase diagram. $B_{C1}$ is temperature dependent, ranging from $B$ $\sim$ 50~mT at $T$ = 10~K to $B$ $\sim$ 10~mT at $T$ = 40~K. 

This strong temperature dependence of $B_{C1}$ is not visible in the contour map of $\chi^\prime$ for FC as shown in Fig. \ref{ZFC_FC}(b). In the Field Cooled case, the helical phase covers a much smaller section of the phase diagram and is suppressed to fields $B_{C1}$ \textless~5~mT in favor of the conical phase that covers a much larger part of the phase diagram. This suppression of the helical phase is consistent with previous neutron scattering experiments\cite{bannenberg2016} and with (ac) susceptibility measurements of Fe$_{1-x}$Co$_{x}$Si with different degrees of Co doping.\cite{bauer2016} This strong temperature and history dependence of $B_{C1}$ has not been observed for MnSi and Cu$_2$OSeO$_3$, but is similar to the doped compounds Mn$_{1-x}$Fe$_x$Si and Mn$_{1-x}$Co$_x$Si,\cite{bauer2010}, where the helical phase also covers a wide section of the magnetic phase diagram for ZFC.

In contrast to $B_{C1}$, the contour plots of $\chi^\prime$ displayed in Figs. \ref{ZFC_FC} (a) and (b) show no history dependence for the borders of the A-phase and the upper magnetic field boundary of the conical phase $B_{C2}$. The A-phase appears as a region with a locally lower $\chi^\prime$ just below $T_C$ in both Figs. \ref{ZFC_FC} (a) and (b) and is bound by the previously defined $B_{A1}$ and $B_{A2}$. Figs. \ref{ZFC_FC}(c) and (d) reveal two clear regions with a non-zero $\chi^\prime$$^\prime$ around $B_{A1}$ and $B_{A2}$. However, neither the lower nor the higher temperature limits of the A-phase are delimited by $\chi^\prime$$^\prime$ implying that the temperature induced transitions to the A-phase are fundamentally different from the field induced ones. This is similar to MnSi\cite{bauer2012} and Cu$_2$OSeO$_3$.\cite{qian2016}

The evolution of $B_{C2}$ as a function of temperature can be described over the whole temperature range of the measurements by the power law $B_{C2} \propto (T_0-T)^{0.39\pm0.04}$ where $T_0$ = 46.1~$\pm$ 0.4~K, i.e. $\sim$$T_C$ + 3~K and is indicated with a dotted line in Figs. \ref{ZFC_FC} (a) and (b). This power law seems to mimic the temperature dependence of a Heisenberg model order parameter. A similar analysis has been performed for Cu$_2$OSeO$_3$ where an exponent of 0.25 was found,\citep{levatic2016} which thus suggest that $B_{C2}$ varies much stronger with temperature for Fe$_{0.7}$Co$_{0.3}$Si than for Cu$_2$OSeO$_3$.

Above $T_C$, the inflection point of $\chi^\prime$ can be observed at $B$ $\sim$45~mT for $T$ = 46~K. At higher temperatures, this inflection point occurs at magnetic fields that increase linearly with increasing temperature extrapolating at zero field to $T_0$. However, for  $T$ $>$ 48~K, the minimum of $d\chi^\prime/dT$ becomes very broad and the deduced inflection points are very inaccurate. For this reason they are not displayed in Fig. 6. The broad maxima in $\chi^\prime$$^\prime$ visible above $T_C$ do not depend on history and will be discussed more extensively in a following section.

\section{\label{sec:level5} Frequency dependence}
\begin{figure}
\begin{center}
\includegraphics[width= .5 \textwidth]{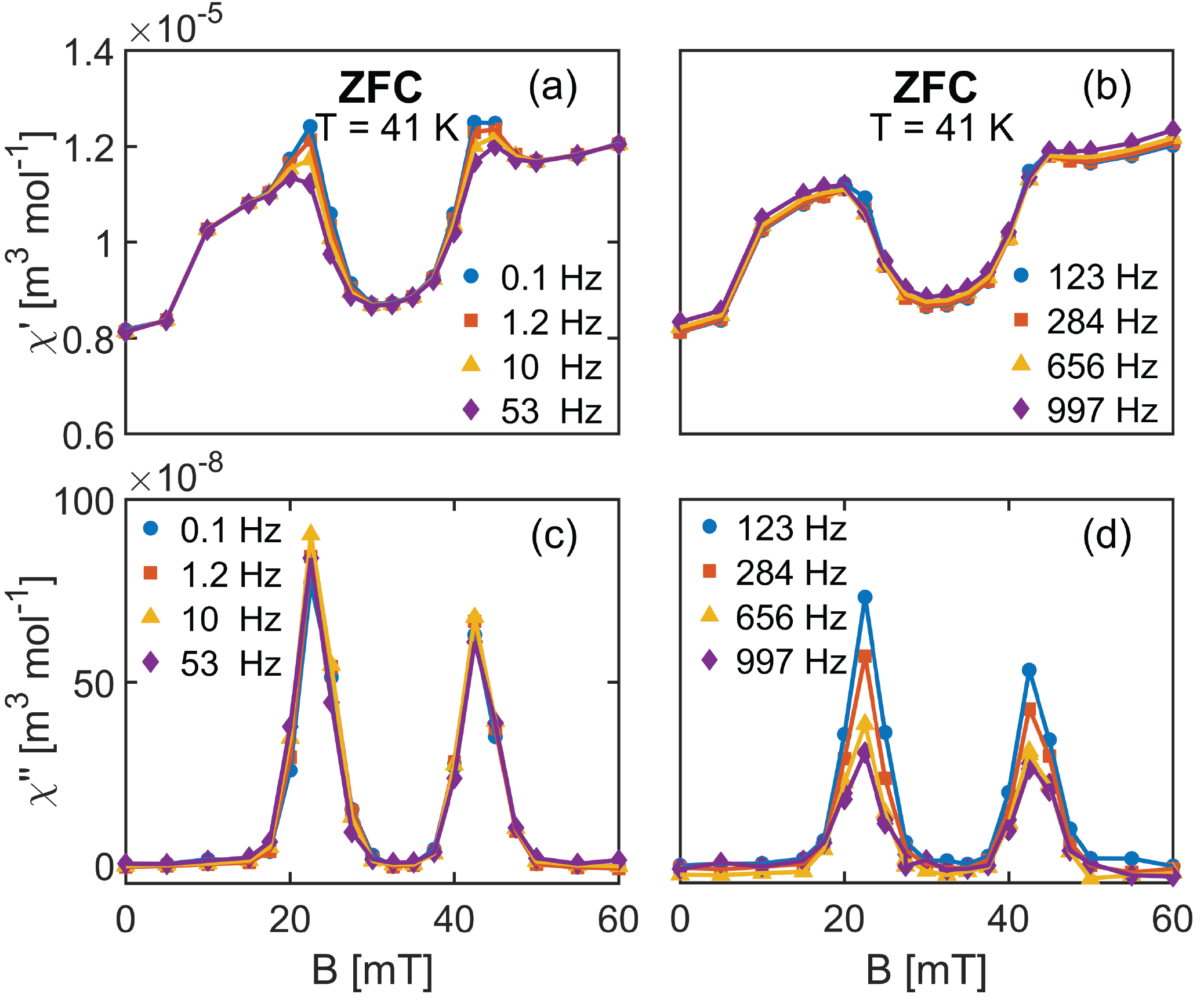}
\caption{Magnetic field dependence of (a) - (b) $\chi^\prime$ and (c) - (d) $\chi^\prime$$^\prime$ as a function of magnetic field for $T$ = 41 K for the frequencies indicated.}
\label{Fig5_freq_41K}
\end{center}
\end{figure}

The results presented at the previous two sections were obtained with an ac drive frequency of 5~Hz only. Nevertheless, the existence of a non-zero $\chi^\prime$$^\prime$ implies a frequency dependence for both $\chi^\prime$ and $\chi^\prime$$^\prime$. This is shown for ZFC in Fig. \ref{Fig5_freq_41K} which displays $\chi^\prime$ and $\chi^\prime$$^\prime$ at $T$ = 41 K as a function of magnetic field for several frequencies. 

Figures \ref{Fig5_freq_41K} (a) and (b) show that this frequency dependence of $\chi^\prime$ is concentrated at the boundary of the A-phase, and that the minimum in the center of the A-phase does not depend on the frequency. The two sharp maxima of $\chi^\prime$ centered at $B$ = 22 mT and $B$ = 42 mT at the boundaries of the A-phase smoothen and soften with increasing frequency. We note that the baseline of $\chi$$^\prime$ (and $\chi$$^{\prime\prime}$) is slightly higher at 1 kHz by $\sim$10$^{-7}$ m$^3 $mol$^{-1}$, which is likely due to the onset of eddy currents.
 
The plots of $\chi$$^\prime$$^\prime$ visible in Figs. \ref{Fig5_freq_41K} (c) and (d) show two sharp maxima close to $B_{A1}$ and $B_{A2}$. These two maxima exhibit a strong frequency dependence as their amplitude decreases significantly for frequencies exceeding 50~Hz resulting in a $\sim$60\% reduction at 1~kHz as compared with 50~Hz. 

\begin{figure}
\begin{center}
\includegraphics[width= .45 \textwidth]{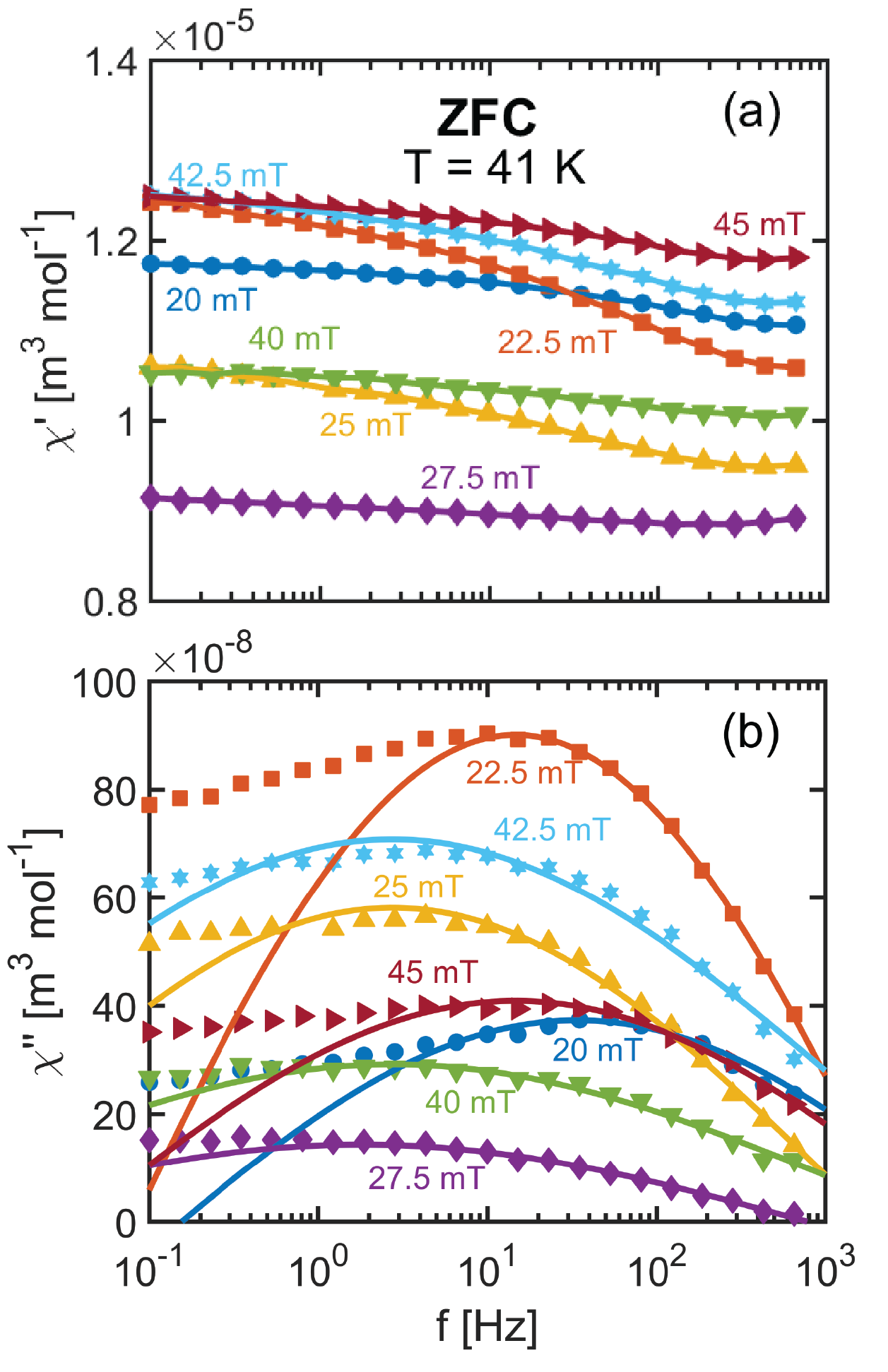}
\caption{Frequency dependence of (a) $\chi^\prime$ and (b) $\chi^\prime$$^\prime$ at various magnetic fields for $T$ = 41~K. The solid lines in panel (b) indicate fits with the relation provided in eq. \ref{eq3}. }
\label{Freqscan}
\end{center}
\end{figure}

%

A more detailed picture of the effect of frequency on the ac susceptibility around the A-phase is given in Fig. \ref{Freqscan}, which displays $\chi^\prime$ and $\chi^\prime$$^\prime$ after ZFC for $T$ = 41~K as a function of frequency for several magnetic fields around $B_{A1}$ and $B_{A2}$. Fig. \ref{Freqscan}(a) shows a relatively weak frequency dependence of $\chi^\prime$ that decreases monotonously with increasing frequency for every magnetic field. In accordance with the behavior shown in Figs. \ref{Fig5_freq_41K} (a) and (b), this dependence is larger for the lower magnetic field limit of the A phase.

Figure \ref{Freqscan} (b) reveals a very broad and asymmetric frequency dependence of $\chi^\prime$$^\prime$. The scans show that the frequencies at which the maxima of $\chi$$^\prime$$^\prime$ occur vary strongly with field. Around $B_{A1}$, the characteristic frequency varies from $\sim$50 Hz at 20~mT to $\sim$10~Hz at 22.5~mT, $\sim$5~Hz at 25~mT, and is in the 0.1~Hz range at 27.5~mT. This indicates that the dynamics become significantly slower towards the center of the A-phase. A complementary behavior is found around $B_{A2}$ where the characteristic frequencies increase substantially with increasing field. The corresponding macroscopic relaxation times indicate extremely slow dynamics in Fe$_{0.7}$Co$_{0.3}$Si that possibly originate from rearrangements of large magnetic volumes. 

Further insights in the relaxation processes behind the frequency dependence is provided by the Cole-Cole formalism that has been modified to include a distribution of relaxation times centered around a characteristic relaxation frequency $f_0$:

\begin{equation}
\chi(\omega) = \chi(\infty) + \frac{\chi(0)-\chi(\infty)}{1+(i\omega\tau_0)^{1-\alpha}},
\label{eq1}
\end{equation}

\noindent where $\omega$ = 2$\pi f$ denotes the angular frequency, $\chi(0)$ and $\chi(\infty)$ the isothermal and adiabatic susceptibility, respectively, $\tau_0$ = 1/2$\pi f_0$ the characteristic relaxation time and $\alpha$ a parameter that provides a measure of the width of the distribution of relaxation frequencies, being zero for a single relaxation process and one for an infinitely broad distribution. A non-zero value of $\alpha$ hence implies a stretched exponential relaxation possibly due to a distributions of energy barriers in a phase-space landscape.\cite{campbell1986} Eq. \ref{eq1} can be decomposed in the in- and out of phase components:\cite{huser1986,dekker1989}

\begin{equation}
\begin{aligned}
\chi(\omega)^\prime = & \chi(\infty)  +  \\ &\frac{(\chi(0)-\chi(\infty))[1+(\omega\tau_0)^{1-\alpha}\sin(\pi\alpha/2)]}{1+2(\omega\tau_0)^{1-\alpha}\sin(\pi\alpha/2)+(\omega\tau_0)^{2(1-\alpha)}},
\end{aligned}
\label{eq2}
\end{equation}

\begin{equation}
\chi(\omega)^{\prime\prime} = \frac{(\chi(0)-\chi(\infty))(\omega\tau_0)^{1-\alpha}\cos(\pi\alpha/2)}{1+2(\omega\tau_0)^{1-\alpha}\sin(\pi\alpha/2)+(\omega\tau_0)^{2(1-\alpha)}}.
\label{eq3}
\end{equation}

\begin{figure*}
\begin{center}
\includegraphics[width= 1\textwidth]{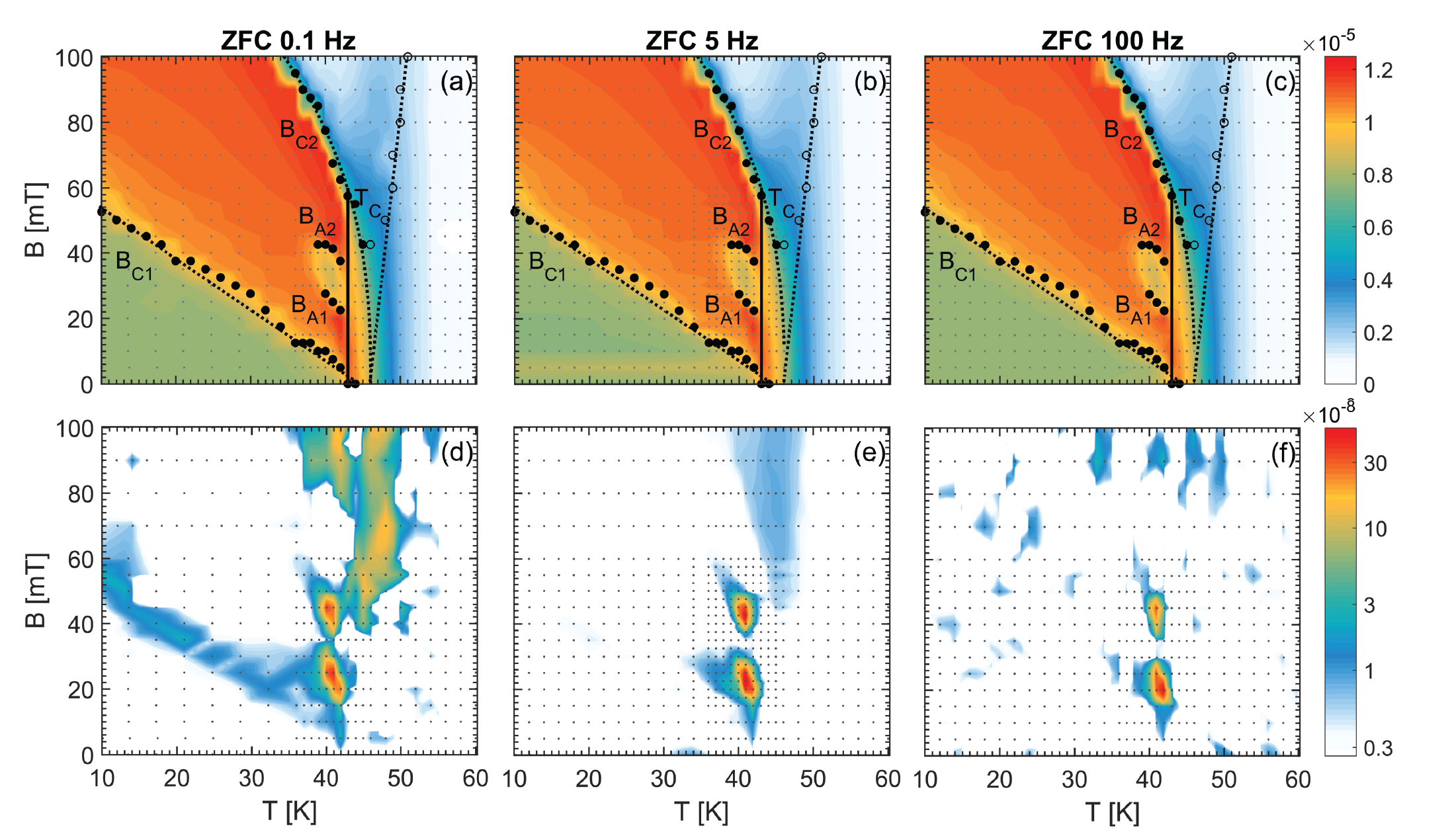}
\caption{Contour plots showing (a) - (c) $\chi^\prime$ and (d) - (f) $\chi^\prime$$^\prime$ in units of m$^3$mol$^{-1}$ for ZFC at $f$ = 0.1, 5 and 100 Hz as a function of temperature and magnetic field. $B_{C1}$, $B_{A1}$, $B_{A2}$ and $B_{C2}$ are defined by the inflection point of $\chi^\prime$ and are indicated with black circles. The dashed line through the $B_{C2}$ points indicate the fitted power law as described in the text in Section IV. The grey dots indicate the points at which a signal has been recorded.$^{42}$}
\label{FreqMaps}
\end{center}
\end{figure*}

The resulting fits for $\chi^\prime$$^\prime$ are displayed in Fig. \ref{Freqscan} (b) and show clear discrepancies at low frequencies and hence fail to accurately describe the entire frequency dependence of the data. The single relaxation process Cole-Cole formalism, even with substantially non-zero values for $\alpha$ required for the fits, fails to describe the frequency dependence of $\chi$$^\prime$ and $\chi$$^\prime$$^\prime$. The behavior observed indicates the existence of several co-existing relaxation processes which might originate from a coexistence of multiple phases as evidenced by neutron scattering.\cite{bannenberg2016}  These relaxation processes occur at low frequencies in Fe$_{0.7}$Co$_{0.3}$Si and give rise to more complicated dynamics around the A-phase than in the non-doped compounds as for example Cu$_2$OSeO$_3$\cite{qian2016} or MnSi.\cite{qian2016MnSi} 

\section{\label{sec:level6} Phase Diagram and Discussion}
An overview of the effect of the frequency on the susceptibility over a wide magnetic field and temperature range is provided by Fig. \ref{FreqMaps} which displays contour plots of $\chi$$^\prime$ and $\chi$$^\prime$$^\prime$ at 0.1~Hz, 5~Hz and 100~Hz after ZFC. The helical-to-conical transition can be clearly identified by the increase in $\chi$$^\prime$ visible at $B_{C1}$ in Figs. \ref{FreqMaps} (a)-(c). On the other hand, Figs. \ref{FreqMaps}(d)-(f) show that around $B_{C1}$ $\chi$$^\prime$$^\prime$ is non-zero only at the lowest frequency of 0.1~Hz. This indicates that the helical-to-conical transition involves very slow dynamics and macroscopic relaxation times of the order of seconds. This is different from Cu$_2$OSeO$_3$ where the characteristic frequencies of the helical-to-conical transition are of the order of $\sim$100 Hz\cite{qian2016}, but could be more similar to MnSi where the characteristic frequency may be well below 10~Hz\cite{bauer2012} or even below 0.1~Hz.\cite{qian2016MnSi} 

In addition, the contour maps of $\chi$$^\prime$$^\prime$ displayed in Figs. \ref{FreqMaps} (d)-(f) reveal a strong frequency dependence of the two regions with non-zero $\chi$$^\prime$$^\prime$ at the lower and higher magnetic field limit of the A-phase around $B_{A1}$ and $B_{A2}$.  Compared with 5 Hz, these regions span a smaller space in the $B$ - $T$ diagram at 100~Hz. This effect is enhanced for the `pocket' of non-zero $\chi$$^\prime$$^\prime$ centered around 45~mT. At 0.1~Hz, the two pockets of non-zero $\chi$$^\prime$$^\prime$ seem to be larger than at 5~Hz.

Moreover, the contour maps of $\chi$$^\prime$$^\prime$  indicate that the non-zero values for $\chi$$^\prime$$^\prime$ visible at 5 Hz above $T_C$ and $B_{C2}$ and centered around $T$ = 46~K   are also visible at 0.1 Hz, where $\chi''$  is about an order of magnitude larger but have disappeared at a frequency of 100~Hz.  A frequency scan at $T$ = 46~K and $B$ = 60~mT reveals a monotonous decrease of $\chi$$^\prime$$^\prime$ with increasing frequency above 0.1~Hz, implying that the characteristic relaxation time associated with this feature exceeds ten seconds.\footnote{At a frequency of 0.1~Hz, the SQUID has difficulties stabilizing the phase in a region from $T$ = 42 to 50~K and fields below 50~mT, which is likely caused by the sample.} This region of non-zero $\chi$$^\prime$$^\prime$ above $B_{C2}$ at low frequencies has also been observed for Cu$_2$OSeO$_3$ (0.8~Hz),\cite{qian2016} MnSi (5~Hz)\cite{qian2016MnSi,tsuruta2016} and the soliton lattice system Cr$_{0.33}$NbS$_{2}$\cite{tsuruta2016} and might be a more generic feature of (cubic) helimagnets. As the origin of this signal remains unclear, we intend to further investigate this feature in the future. 

The occurrence of metastable skyrmion lattice correlations at low temperatures under field cooling with cooling rates of $\sim$0.1 Kmin$^{-1}$ for FC\footnote{The cooling rate reported here for FC refers to the average cooling rate between $T$ = 40~K and 30~K.} and $\sim$10 Kmin$^{-1}$ for Fast FC are exceptional and might also be related to the slow dynamics of the system probed by the frequency scans. The unwinding or decay of the skyrmion lattice correlations in Fe$_{0.7}$Co$_{0.3}$Si is considerably slower than for MnSi, where much higher cooling rates of $\sim$700~Ks$^{-1}$ are required to freeze the skyrmion lattice correlations below the A-phase. \cite{milde2013,oike2016,bannenberg2016,bauer2016} In other words, the slow decay of skyrmion lattice correlations is likely related to the high degree of chemical disorder in the system and can be (partially) prevented with conventional cooling rates. 

\section{\label{sec:level7} Conclusion}
The systematic study of the AC susceptibility of Fe$_{0.7}$Co$_{0.3}$Si presented above confirms the dependence of the magnetic phase diagram on the magnetic history and the applied cooling rates reported in previous studies. The transitions between the helical, conical and A-phase can be derived from $\chi$$^\prime$ and show that for Zero Field Cooling  the helical phase covers a wide section of the phase diagram with a critical field depending on temperature whereas no temperature dependence is found for Field Cooling. 

The weak frequency dependence of $\chi$$^\prime$ is in sharp contrast with the strong frequency dependence of $\chi$$^\prime$$^\prime$. Around the A-phase, this is an asymmetric and broad frequency dependence arising from several co-existing relaxation processes with characteristic relaxation times ranging from tens of milliseconds to several seconds. In addition, a $\chi$$^\prime$$^\prime$ signal is found at the helical-to-conical transition but only for the lowest frequency applied of 0.1~Hz. Moreover, a non-zero $\chi$$^\prime$$^\prime$ is observed above $T_C$ at low frequencies and in a wide region of the phase diagram. Albeit the numerous similarities with other chiral systems, the pronounced history and cooling rate dependence of the magnetic phase diagram on the magnetic history as well as the asymmetric frequency dependence and slow dynamics are special to Fe$_{0.7}$Co$_{0.3}$Si and suggest more complicated physical phenomena than in Cu$_2$OSeO$_3$ and MnSi.

\begin{acknowledgments}
The authors wish to thank G.R. Blake for the Laue x-ray diffraction measurements and the alignment of the sample. The work of LB is financially supported by The Netherlands Organization for Scientific Research through the NWO Groot project LARMOR 721.012.102.  
\end{acknowledgments}

\bibliography{FeCoSiSQUID}

\begin{thebibliography}{45}%
\makeatletter
\providecommand \@ifxundefined [1]{%
 \@ifx{#1\undefined}
}%
\providecommand \@ifnum [1]{%
 \ifnum #1\expandafter \@firstoftwo
 \else \expandafter \@secondoftwo
 \fi
}%
\providecommand \@ifx [1]{%
 \ifx #1\expandafter \@firstoftwo
 \else \expandafter \@secondoftwo
 \fi
}%
\providecommand \natexlab [1]{#1}%
\providecommand \enquote  [1]{``#1''}%
\providecommand \bibnamefont  [1]{#1}%
\providecommand \bibfnamefont [1]{#1}%
\providecommand \citenamefont [1]{#1}%
\providecommand \href@noop [0]{\@secondoftwo}%
\providecommand \href [0]{\begingroup \@sanitize@url \@href}%
\providecommand \@href[1]{\@@startlink{#1}\@@href}%
\providecommand \@@href[1]{\endgroup#1\@@endlink}%
\providecommand \@sanitize@url [0]{\catcode `\\12\catcode `\$12\catcode
  `\&12\catcode `\#12\catcode `\^12\catcode `\_12\catcode `\%12\relax}%
\providecommand \@@startlink[1]{}%
\providecommand \@@endlink[0]{}%
\providecommand \url  [0]{\begingroup\@sanitize@url \@url }%
\providecommand \@url [1]{\endgroup\@href {#1}{\urlprefix }}%
\providecommand \urlprefix  [0]{URL }%
\providecommand \Eprint [0]{\href }%
\providecommand \doibase [0]{http://dx.doi.org/}%
\providecommand \selectlanguage [0]{\@gobble}%
\providecommand \bibinfo  [0]{\@secondoftwo}%
\providecommand \bibfield  [0]{\@secondoftwo}%
\providecommand \translation [1]{[#1]}%
\providecommand \BibitemOpen [0]{}%
\providecommand \bibitemStop [0]{}%
\providecommand \bibitemNoStop [0]{.\EOS\space}%
\providecommand \EOS [0]{\spacefactor3000\relax}%
\providecommand \BibitemShut  [1]{\csname bibitem#1\endcsname}%
\let\auto@bib@innerbib\@empty
\bibitem [{\citenamefont {M{\"u}hlbauer}\ \emph {et~al.}(2009)\citenamefont
  {M{\"u}hlbauer}, \citenamefont {Binz}, \citenamefont {Jonietz}, \citenamefont
  {Pfleiderer}, \citenamefont {Rosch}, \citenamefont {Neubauer}, \citenamefont
  {Georgii},\ and\ \citenamefont {B{\"o}ni}}]{muhlbauer2009}%
  \BibitemOpen
  \bibfield  {author} {\bibinfo {author} {\bibfnamefont {S.}~\bibnamefont
  {M{\"u}hlbauer}}, \bibinfo {author} {\bibfnamefont {B.}~\bibnamefont {Binz}},
  \bibinfo {author} {\bibfnamefont {F.}~\bibnamefont {Jonietz}}, \bibinfo
  {author} {\bibfnamefont {C.}~\bibnamefont {Pfleiderer}}, \bibinfo {author}
  {\bibfnamefont {A.}~\bibnamefont {Rosch}}, \bibinfo {author} {\bibfnamefont
  {A.}~\bibnamefont {Neubauer}}, \bibinfo {author} {\bibfnamefont
  {R.}~\bibnamefont {Georgii}}, \ and\ \bibinfo {author} {\bibfnamefont
  {P.}~\bibnamefont {B{\"o}ni}},\ }\href@noop {} {\bibfield  {journal}
  {\bibinfo  {journal} {Science}\ }\textbf {\bibinfo {volume} {323}},\ \bibinfo
  {pages} {915} (\bibinfo {year} {2009})}\BibitemShut {NoStop}%
\bibitem [{\citenamefont {Yu}\ \emph {et~al.}(2010{\natexlab{a}})\citenamefont
  {Yu}, \citenamefont {Kanazawa}, \citenamefont {Onose}, \citenamefont
  {Kimoto}, \citenamefont {Zhang}, \citenamefont {Ishiwata}, \citenamefont
  {Matsui},\ and\ \citenamefont {Tokura}}]{FeGe_Lorenz_TEM}%
  \BibitemOpen
  \bibfield  {author} {\bibinfo {author} {\bibfnamefont {X.~Z.}\ \bibnamefont
  {Yu}}, \bibinfo {author} {\bibfnamefont {N.}~\bibnamefont {Kanazawa}},
  \bibinfo {author} {\bibfnamefont {Y.}~\bibnamefont {Onose}}, \bibinfo
  {author} {\bibfnamefont {K.}~\bibnamefont {Kimoto}}, \bibinfo {author}
  {\bibfnamefont {W.~Z.}\ \bibnamefont {Zhang}}, \bibinfo {author}
  {\bibfnamefont {S.}~\bibnamefont {Ishiwata}}, \bibinfo {author}
  {\bibfnamefont {Y.}~\bibnamefont {Matsui}}, \ and\ \bibinfo {author}
  {\bibfnamefont {Y.}~\bibnamefont {Tokura}},\ }\href@noop {} {\bibfield
  {journal} {\bibinfo  {journal} {Nature Materials}\ }\textbf {\bibinfo
  {volume} {10}},\ \bibinfo {pages} {106} (\bibinfo {year}
  {2010}{\natexlab{a}})}\BibitemShut {NoStop}%
\bibitem [{\citenamefont {Moskvin}\ \emph {et~al.}(2013)\citenamefont
  {Moskvin}, \citenamefont {Grigoriev}, \citenamefont {Dyadkin}, \citenamefont
  {Eckerlebe}, \citenamefont {Baenitz}, \citenamefont {Schmidt},\ and\
  \citenamefont {Wilhelm}}]{moskvin2013}%
  \BibitemOpen
  \bibfield  {author} {\bibinfo {author} {\bibfnamefont {E.}~\bibnamefont
  {Moskvin}}, \bibinfo {author} {\bibfnamefont {S.}~\bibnamefont {Grigoriev}},
  \bibinfo {author} {\bibfnamefont {V.}~\bibnamefont {Dyadkin}}, \bibinfo
  {author} {\bibfnamefont {H.}~\bibnamefont {Eckerlebe}}, \bibinfo {author}
  {\bibfnamefont {M.}~\bibnamefont {Baenitz}}, \bibinfo {author} {\bibfnamefont
  {M.}~\bibnamefont {Schmidt}}, \ and\ \bibinfo {author} {\bibfnamefont
  {H.}~\bibnamefont {Wilhelm}},\ }\href@noop {} {\bibfield  {journal} {\bibinfo
   {journal} {Physical Review Letters}\ }\textbf {\bibinfo {volume} {110}},\
  \bibinfo {pages} {077207} (\bibinfo {year} {2013})}\BibitemShut {NoStop}%
\bibitem [{\citenamefont {Seki}\ \emph
  {et~al.}(2012{\natexlab{a}})\citenamefont {Seki}, \citenamefont {Yu},
  \citenamefont {Ishiwata},\ and\ \citenamefont
  {Tokura}}]{seki2012observation}%
  \BibitemOpen
  \bibfield  {author} {\bibinfo {author} {\bibfnamefont {S.}~\bibnamefont
  {Seki}}, \bibinfo {author} {\bibfnamefont {X.~Z.}\ \bibnamefont {Yu}},
  \bibinfo {author} {\bibfnamefont {S.}~\bibnamefont {Ishiwata}}, \ and\
  \bibinfo {author} {\bibfnamefont {Y.}~\bibnamefont {Tokura}},\ }\href@noop {}
  {\bibfield  {journal} {\bibinfo  {journal} {Science}\ }\textbf {\bibinfo
  {volume} {336}},\ \bibinfo {pages} {198} (\bibinfo {year}
  {2012}{\natexlab{a}})}\BibitemShut {NoStop}%
\bibitem [{\citenamefont {Seki}\ \emph
  {et~al.}(2012{\natexlab{b}})\citenamefont {Seki}, \citenamefont {Kim},
  \citenamefont {Inosov}, \citenamefont {Georgii}, \citenamefont {Keimer},
  \citenamefont {Ishiwata},\ and\ \citenamefont {Tokura}}]{seki2012formation}%
  \BibitemOpen
  \bibfield  {author} {\bibinfo {author} {\bibfnamefont {S.}~\bibnamefont
  {Seki}}, \bibinfo {author} {\bibfnamefont {J.-H.}\ \bibnamefont {Kim}},
  \bibinfo {author} {\bibfnamefont {D.~S.}\ \bibnamefont {Inosov}}, \bibinfo
  {author} {\bibfnamefont {R.}~\bibnamefont {Georgii}}, \bibinfo {author}
  {\bibfnamefont {B.}~\bibnamefont {Keimer}}, \bibinfo {author} {\bibfnamefont
  {S.}~\bibnamefont {Ishiwata}}, \ and\ \bibinfo {author} {\bibfnamefont
  {Y.}~\bibnamefont {Tokura}},\ }\href@noop {} {\bibfield  {journal} {\bibinfo
  {journal} {Physical Review B}\ }\textbf {\bibinfo {volume} {85}},\ \bibinfo
  {pages} {220406} (\bibinfo {year} {2012}{\natexlab{b}})}\BibitemShut
  {NoStop}%
\bibitem [{\citenamefont {Adams}\ \emph {et~al.}(2012)\citenamefont {Adams},
  \citenamefont {Chacon}, \citenamefont {Wagner}, \citenamefont {Bauer},
  \citenamefont {Brandl}, \citenamefont {Pedersen}, \citenamefont {Berger},
  \citenamefont {Lemmens},\ and\ \citenamefont {Pfleiderer}}]{adams2012long}%
  \BibitemOpen
  \bibfield  {author} {\bibinfo {author} {\bibfnamefont {T.}~\bibnamefont
  {Adams}}, \bibinfo {author} {\bibfnamefont {A.}~\bibnamefont {Chacon}},
  \bibinfo {author} {\bibfnamefont {M.}~\bibnamefont {Wagner}}, \bibinfo
  {author} {\bibfnamefont {A.}~\bibnamefont {Bauer}}, \bibinfo {author}
  {\bibfnamefont {G.}~\bibnamefont {Brandl}}, \bibinfo {author} {\bibfnamefont
  {B.}~\bibnamefont {Pedersen}}, \bibinfo {author} {\bibfnamefont
  {H.}~\bibnamefont {Berger}}, \bibinfo {author} {\bibfnamefont
  {P.}~\bibnamefont {Lemmens}}, \ and\ \bibinfo {author} {\bibfnamefont
  {C.}~\bibnamefont {Pfleiderer}},\ }\href@noop {} {\bibfield  {journal}
  {\bibinfo  {journal} {Physical Review Letters}\ }\textbf {\bibinfo {volume}
  {108}},\ \bibinfo {pages} {237204} (\bibinfo {year} {2012})}\BibitemShut
  {NoStop}%
\bibitem [{\citenamefont {Yu}\ \emph {et~al.}(2010{\natexlab{b}})\citenamefont
  {Yu}, \citenamefont {Onose}, \citenamefont {Kanazawa}, \citenamefont {Park},
  \citenamefont {Han}, \citenamefont {Matsui}, \citenamefont {Nagaosa},\ and\
  \citenamefont {Tokura}}]{yu2010}%
  \BibitemOpen
  \bibfield  {author} {\bibinfo {author} {\bibfnamefont {X.~Z.}\ \bibnamefont
  {Yu}}, \bibinfo {author} {\bibfnamefont {Y.}~\bibnamefont {Onose}}, \bibinfo
  {author} {\bibfnamefont {N.}~\bibnamefont {Kanazawa}}, \bibinfo {author}
  {\bibfnamefont {J.~H.}\ \bibnamefont {Park}}, \bibinfo {author}
  {\bibfnamefont {J.~H.}\ \bibnamefont {Han}}, \bibinfo {author} {\bibfnamefont
  {Y.}~\bibnamefont {Matsui}}, \bibinfo {author} {\bibfnamefont
  {N.}~\bibnamefont {Nagaosa}}, \ and\ \bibinfo {author} {\bibfnamefont
  {Y.}~\bibnamefont {Tokura}},\ }\href@noop {} {\bibfield  {journal} {\bibinfo
  {journal} {Nature}\ }\textbf {\bibinfo {volume} {465}},\ \bibinfo {pages}
  {901} (\bibinfo {year} {2010}{\natexlab{b}})}\BibitemShut {NoStop}%
\bibitem [{\citenamefont {M{\"u}nzer}\ \emph {et~al.}(2010)\citenamefont
  {M{\"u}nzer}, \citenamefont {Neubauer}, \citenamefont {Adams}, \citenamefont
  {M{\"u}hlbauer}, \citenamefont {Franz}, \citenamefont {Jonietz},
  \citenamefont {Georgii}, \citenamefont {B{\"o}ni}, \citenamefont {Pedersen},
  \citenamefont {Schmidt} \emph {et~al.}}]{munzer2010}%
  \BibitemOpen
  \bibfield  {author} {\bibinfo {author} {\bibfnamefont {W.}~\bibnamefont
  {M{\"u}nzer}}, \bibinfo {author} {\bibfnamefont {A.}~\bibnamefont
  {Neubauer}}, \bibinfo {author} {\bibfnamefont {T.}~\bibnamefont {Adams}},
  \bibinfo {author} {\bibfnamefont {S.}~\bibnamefont {M{\"u}hlbauer}}, \bibinfo
  {author} {\bibfnamefont {C.}~\bibnamefont {Franz}}, \bibinfo {author}
  {\bibfnamefont {F.}~\bibnamefont {Jonietz}}, \bibinfo {author} {\bibfnamefont
  {R.}~\bibnamefont {Georgii}}, \bibinfo {author} {\bibfnamefont
  {P.}~\bibnamefont {B{\"o}ni}}, \bibinfo {author} {\bibfnamefont
  {B.}~\bibnamefont {Pedersen}}, \bibinfo {author} {\bibfnamefont
  {M.}~\bibnamefont {Schmidt}},  \emph {et~al.},\ }\href@noop {} {\bibfield
  {journal} {\bibinfo  {journal} {Physical Review B}\ }\textbf {\bibinfo
  {volume} {81}},\ \bibinfo {pages} {041203} (\bibinfo {year}
  {2010})}\BibitemShut {NoStop}%
\bibitem [{\citenamefont {Dzyaloshinski}(1958)}]{D}%
  \BibitemOpen
  \bibfield  {author} {\bibinfo {author} {\bibfnamefont {I.~E.}\ \bibnamefont
  {Dzyaloshinski}},\ }\href@noop {} {\bibfield  {journal} {\bibinfo  {journal}
  {J. Phys. Chem. Solids}\ }\textbf {\bibinfo {volume} {4}},\ \bibinfo {pages}
  {241} (\bibinfo {year} {1958})}\BibitemShut {NoStop}%
\bibitem [{\citenamefont {Moriya}(1960)}]{M}%
  \BibitemOpen
  \bibfield  {author} {\bibinfo {author} {\bibfnamefont {T.}~\bibnamefont
  {Moriya}},\ }\href@noop {} {\bibfield  {journal} {\bibinfo  {journal} {Phys.
  Rev}\ }\textbf {\bibinfo {volume} {120}},\ \bibinfo {pages} {91} (\bibinfo
  {year} {1960})}\BibitemShut {NoStop}%
\bibitem [{\citenamefont {Bogdanov}\ and\ \citenamefont
  {Yablonskii}(1989)}]{bogdanov1989}%
  \BibitemOpen
  \bibfield  {author} {\bibinfo {author} {\bibfnamefont {A.~N.}\ \bibnamefont
  {Bogdanov}}\ and\ \bibinfo {author} {\bibfnamefont {D.~A.}\ \bibnamefont
  {Yablonskii}},\ }\href@noop {} {\bibfield  {journal} {\bibinfo  {journal}
  {Sov. Phys. JETP}\ }\textbf {\bibinfo {volume} {68}},\ \bibinfo {pages} {101}
  (\bibinfo {year} {1989})}\BibitemShut {NoStop}%
\bibitem [{\citenamefont {Bogdanov}\ and\ \citenamefont
  {Hubert}(1994)}]{bogdanov1994}%
  \BibitemOpen
  \bibfield  {author} {\bibinfo {author} {\bibfnamefont {A.~N.}\ \bibnamefont
  {Bogdanov}}\ and\ \bibinfo {author} {\bibfnamefont {A.}~\bibnamefont
  {Hubert}},\ }\href@noop {} {\bibfield  {journal} {\bibinfo  {journal}
  {Journal of magnetism and magnetic materials}\ }\textbf {\bibinfo {volume}
  {138}},\ \bibinfo {pages} {255} (\bibinfo {year} {1994})}\BibitemShut
  {NoStop}%
\bibitem [{\citenamefont {R{\"o}{\ss}ler}\ \emph {et~al.}(2006)\citenamefont
  {R{\"o}{\ss}ler}, \citenamefont {Bogdanov},\ and\ \citenamefont
  {Pfleiderer}}]{Rossler2006}%
  \BibitemOpen
  \bibfield  {author} {\bibinfo {author} {\bibfnamefont {U.~K.}\ \bibnamefont
  {R{\"o}{\ss}ler}}, \bibinfo {author} {\bibfnamefont {A.~N.}\ \bibnamefont
  {Bogdanov}}, \ and\ \bibinfo {author} {\bibfnamefont {C.}~\bibnamefont
  {Pfleiderer}},\ }\href@noop {} {\bibfield  {journal} {\bibinfo  {journal}
  {Nature}\ }\textbf {\bibinfo {volume} {442}},\ \bibinfo {pages} {797}
  (\bibinfo {year} {2006})}\BibitemShut {NoStop}%
\bibitem [{\citenamefont {Nagaosa}\ and\ \citenamefont
  {Tokura}(2013)}]{nagaosa2013}%
  \BibitemOpen
  \bibfield  {author} {\bibinfo {author} {\bibfnamefont {N.}~\bibnamefont
  {Nagaosa}}\ and\ \bibinfo {author} {\bibfnamefont {Y.}~\bibnamefont
  {Tokura}},\ }\href@noop {} {\bibfield  {journal} {\bibinfo  {journal} {Nature
  Nanotechnology}\ }\textbf {\bibinfo {volume} {8}},\ \bibinfo {pages} {899}
  (\bibinfo {year} {2013})}\BibitemShut {NoStop}%
\bibitem [{\citenamefont {Fert}\ \emph {et~al.}(2013)\citenamefont {Fert},
  \citenamefont {Cros},\ and\ \citenamefont {Sampaio}}]{fert2013}%
  \BibitemOpen
  \bibfield  {author} {\bibinfo {author} {\bibfnamefont {A.}~\bibnamefont
  {Fert}}, \bibinfo {author} {\bibfnamefont {V.}~\bibnamefont {Cros}}, \ and\
  \bibinfo {author} {\bibfnamefont {J.}~\bibnamefont {Sampaio}},\ }\href@noop
  {} {\bibfield  {journal} {\bibinfo  {journal} {Nature nanotechnology}\
  }\textbf {\bibinfo {volume} {8}},\ \bibinfo {pages} {152} (\bibinfo {year}
  {2013})}\BibitemShut {NoStop}%
\bibitem [{\citenamefont {Romming}\ \emph {et~al.}(2013)\citenamefont
  {Romming}, \citenamefont {Hanneken}, \citenamefont {Menzel}, \citenamefont
  {Bickel}, \citenamefont {Wolter}, \citenamefont {von Bergmann}, \citenamefont
  {Kubetzka},\ and\ \citenamefont {Wiesendanger}}]{romming2013}%
  \BibitemOpen
  \bibfield  {author} {\bibinfo {author} {\bibfnamefont {N.}~\bibnamefont
  {Romming}}, \bibinfo {author} {\bibfnamefont {C.}~\bibnamefont {Hanneken}},
  \bibinfo {author} {\bibfnamefont {M.}~\bibnamefont {Menzel}}, \bibinfo
  {author} {\bibfnamefont {J.~E.}\ \bibnamefont {Bickel}}, \bibinfo {author}
  {\bibfnamefont {B.}~\bibnamefont {Wolter}}, \bibinfo {author} {\bibfnamefont
  {K.}~\bibnamefont {von Bergmann}}, \bibinfo {author} {\bibfnamefont
  {A.}~\bibnamefont {Kubetzka}}, \ and\ \bibinfo {author} {\bibfnamefont
  {R.}~\bibnamefont {Wiesendanger}},\ }\href@noop {} {\bibfield  {journal}
  {\bibinfo  {journal} {Science}\ }\textbf {\bibinfo {volume} {341}},\ \bibinfo
  {pages} {636} (\bibinfo {year} {2013})}\BibitemShut {NoStop}%
\bibitem [{\citenamefont {Oike}\ \emph {et~al.}(2016)\citenamefont {Oike},
  \citenamefont {Kikkawa}, \citenamefont {Kanazawa}, \citenamefont {Taguchi},
  \citenamefont {Kawasaki}, \citenamefont {Tokura},\ and\ \citenamefont
  {Kagawa}}]{oike2016}%
  \BibitemOpen
  \bibfield  {author} {\bibinfo {author} {\bibfnamefont {H.}~\bibnamefont
  {Oike}}, \bibinfo {author} {\bibfnamefont {A.}~\bibnamefont {Kikkawa}},
  \bibinfo {author} {\bibfnamefont {N.}~\bibnamefont {Kanazawa}}, \bibinfo
  {author} {\bibfnamefont {Y.}~\bibnamefont {Taguchi}}, \bibinfo {author}
  {\bibfnamefont {M.}~\bibnamefont {Kawasaki}}, \bibinfo {author}
  {\bibfnamefont {Y.}~\bibnamefont {Tokura}}, \ and\ \bibinfo {author}
  {\bibfnamefont {F.}~\bibnamefont {Kagawa}},\ }\href@noop {} {\bibfield
  {journal} {\bibinfo  {journal} {Nature Physics}\ }\textbf {\bibinfo {volume}
  {12}},\ \bibinfo {pages} {62} (\bibinfo {year} {2016})}\BibitemShut {NoStop}%
\bibitem [{\citenamefont {Beille}\ \emph {et~al.}(1981)\citenamefont {Beille},
  \citenamefont {Voiron}, \citenamefont {Towfiq}, \citenamefont {Roth},\ and\
  \citenamefont {Zhang}}]{beille1981}%
  \BibitemOpen
  \bibfield  {author} {\bibinfo {author} {\bibfnamefont {J.}~\bibnamefont
  {Beille}}, \bibinfo {author} {\bibfnamefont {J.}~\bibnamefont {Voiron}},
  \bibinfo {author} {\bibfnamefont {F.}~\bibnamefont {Towfiq}}, \bibinfo
  {author} {\bibfnamefont {M.}~\bibnamefont {Roth}}, \ and\ \bibinfo {author}
  {\bibfnamefont {Z.~Y.}\ \bibnamefont {Zhang}},\ }\href@noop {} {\bibfield
  {journal} {\bibinfo  {journal} {Journal of Physics F: Metal Physics}\
  }\textbf {\bibinfo {volume} {11}},\ \bibinfo {pages} {2153} (\bibinfo {year}
  {1981})}\BibitemShut {NoStop}%
\bibitem [{\citenamefont {Beille}\ \emph {et~al.}(1983)\citenamefont {Beille},
  \citenamefont {Voiron},\ and\ \citenamefont {Roth}}]{beille1983}%
  \BibitemOpen
  \bibfield  {author} {\bibinfo {author} {\bibfnamefont {J.}~\bibnamefont
  {Beille}}, \bibinfo {author} {\bibfnamefont {J.}~\bibnamefont {Voiron}}, \
  and\ \bibinfo {author} {\bibfnamefont {M.}~\bibnamefont {Roth}},\ }\href@noop
  {} {\bibfield  {journal} {\bibinfo  {journal} {Solid state communications}\
  }\textbf {\bibinfo {volume} {47}},\ \bibinfo {pages} {399} (\bibinfo {year}
  {1983})}\BibitemShut {NoStop}%
\bibitem [{\citenamefont {Motokawa}\ \emph {et~al.}(1987)\citenamefont
  {Motokawa}, \citenamefont {Kawarazaki}, \citenamefont {Nojiri},\ and\
  \citenamefont {Inoue}}]{motokawa1987}%
  \BibitemOpen
  \bibfield  {author} {\bibinfo {author} {\bibfnamefont {M.}~\bibnamefont
  {Motokawa}}, \bibinfo {author} {\bibfnamefont {S.}~\bibnamefont
  {Kawarazaki}}, \bibinfo {author} {\bibfnamefont {H.}~\bibnamefont {Nojiri}},
  \ and\ \bibinfo {author} {\bibfnamefont {T.}~\bibnamefont {Inoue}},\
  }\href@noop {} {\bibfield  {journal} {\bibinfo  {journal} {Journal of
  Magnetism and Magnetic Materials}\ }\textbf {\bibinfo {volume} {70}},\
  \bibinfo {pages} {245} (\bibinfo {year} {1987})}\BibitemShut {NoStop}%
\bibitem [{\citenamefont {Onose}\ \emph {et~al.}(2005)\citenamefont {Onose},
  \citenamefont {Takeshita}, \citenamefont {Terakura}, \citenamefont {Takagi},\
  and\ \citenamefont {Tokura}}]{onose2005}%
  \BibitemOpen
  \bibfield  {author} {\bibinfo {author} {\bibfnamefont {Y.}~\bibnamefont
  {Onose}}, \bibinfo {author} {\bibfnamefont {N.}~\bibnamefont {Takeshita}},
  \bibinfo {author} {\bibfnamefont {C.}~\bibnamefont {Terakura}}, \bibinfo
  {author} {\bibfnamefont {H.}~\bibnamefont {Takagi}}, \ and\ \bibinfo {author}
  {\bibfnamefont {Y.}~\bibnamefont {Tokura}},\ }\href@noop {} {\bibfield
  {journal} {\bibinfo  {journal} {Physical Review B}\ }\textbf {\bibinfo
  {volume} {72}},\ \bibinfo {pages} {224431} (\bibinfo {year}
  {2005})}\BibitemShut {NoStop}%
\bibitem [{\citenamefont {Siegfried}\ \emph {et~al.}(2015)\citenamefont
  {Siegfried}, \citenamefont {Altynbaev}, \citenamefont {Chubova},
  \citenamefont {Dyadkin}, \citenamefont {Chernyshov}, \citenamefont {Moskvin},
  \citenamefont {Menzel}, \citenamefont {Heinemann}, \citenamefont {Schreyer},\
  and\ \citenamefont {Grigoriev}}]{siegfried2015}%
  \BibitemOpen
  \bibfield  {author} {\bibinfo {author} {\bibfnamefont {S.-A.}\ \bibnamefont
  {Siegfried}}, \bibinfo {author} {\bibfnamefont {E.~V.}\ \bibnamefont
  {Altynbaev}}, \bibinfo {author} {\bibfnamefont {N.~M.}\ \bibnamefont
  {Chubova}}, \bibinfo {author} {\bibfnamefont {V.}~\bibnamefont {Dyadkin}},
  \bibinfo {author} {\bibfnamefont {D.}~\bibnamefont {Chernyshov}}, \bibinfo
  {author} {\bibfnamefont {E.~V.}\ \bibnamefont {Moskvin}}, \bibinfo {author}
  {\bibfnamefont {D.}~\bibnamefont {Menzel}}, \bibinfo {author} {\bibfnamefont
  {A.}~\bibnamefont {Heinemann}}, \bibinfo {author} {\bibfnamefont
  {A.}~\bibnamefont {Schreyer}}, \ and\ \bibinfo {author} {\bibfnamefont
  {S.~V.}\ \bibnamefont {Grigoriev}},\ }\href@noop {} {\bibfield  {journal}
  {\bibinfo  {journal} {Physical Review B}\ }\textbf {\bibinfo {volume} {91}},\
  \bibinfo {pages} {184406} (\bibinfo {year} {2015})}\BibitemShut {NoStop}%
\bibitem [{\citenamefont {Grigoriev}\ \emph {et~al.}(2009)\citenamefont
  {Grigoriev}, \citenamefont {Chernyshov}, \citenamefont {Dyadkin},
  \citenamefont {Dmitriev}, \citenamefont {Maleyev}, \citenamefont {Moskvin},
  \citenamefont {Menzel}, \citenamefont {Schoenes},\ and\ \citenamefont
  {Eckerlebe}}]{grigoriev2009}%
  \BibitemOpen
  \bibfield  {author} {\bibinfo {author} {\bibfnamefont {S.~V.}\ \bibnamefont
  {Grigoriev}}, \bibinfo {author} {\bibfnamefont {D.}~\bibnamefont
  {Chernyshov}}, \bibinfo {author} {\bibfnamefont {V.~A.}\ \bibnamefont
  {Dyadkin}}, \bibinfo {author} {\bibfnamefont {V.}~\bibnamefont {Dmitriev}},
  \bibinfo {author} {\bibfnamefont {S.~V.}\ \bibnamefont {Maleyev}}, \bibinfo
  {author} {\bibfnamefont {E.~V.}\ \bibnamefont {Moskvin}}, \bibinfo {author}
  {\bibfnamefont {D.}~\bibnamefont {Menzel}}, \bibinfo {author} {\bibfnamefont
  {J.}~\bibnamefont {Schoenes}}, \ and\ \bibinfo {author} {\bibfnamefont
  {H.}~\bibnamefont {Eckerlebe}},\ }\href@noop {} {\bibfield  {journal}
  {\bibinfo  {journal} {Physical Review Letters}\ }\textbf {\bibinfo {volume}
  {102}},\ \bibinfo {pages} {037204} (\bibinfo {year} {2009})}\BibitemShut
  {NoStop}%
\bibitem [{\citenamefont {Bauer}\ \emph {et~al.}(2016)\citenamefont {Bauer},
  \citenamefont {Garst},\ and\ \citenamefont {Pfleiderer}}]{bauer2016}%
  \BibitemOpen
  \bibfield  {author} {\bibinfo {author} {\bibfnamefont {A.}~\bibnamefont
  {Bauer}}, \bibinfo {author} {\bibfnamefont {M.}~\bibnamefont {Garst}}, \ and\
  \bibinfo {author} {\bibfnamefont {C.}~\bibnamefont {Pfleiderer}},\
  }\href@noop {} {\bibfield  {journal} {\bibinfo  {journal} {Physical Review
  B}\ }\textbf {\bibinfo {volume} {93}},\ \bibinfo {pages} {235144} (\bibinfo
  {year} {2016})}\BibitemShut {NoStop}%
\bibitem [{\citenamefont {Bannenberg}\ \emph {et~al.}(2016)\citenamefont
  {Bannenberg}, \citenamefont {Kakurai}, \citenamefont {Qian}, \citenamefont
  {{Leli\`{e}vre-Berna}}, \citenamefont {Dewhurst}, \citenamefont {Onose},
  \citenamefont {Endoh}, \citenamefont {Tokura},\ and\ \citenamefont
  {Pappas}}]{bannenberg2016}%
  \BibitemOpen
  \bibfield  {author} {\bibinfo {author} {\bibfnamefont {L.~J.}\ \bibnamefont
  {Bannenberg}}, \bibinfo {author} {\bibfnamefont {K.}~\bibnamefont {Kakurai}},
  \bibinfo {author} {\bibfnamefont {F.}~\bibnamefont {Qian}}, \bibinfo {author}
  {\bibfnamefont {E.}~\bibnamefont {{Leli\`{e}vre-Berna}}}, \bibinfo {author}
  {\bibfnamefont {C.~D.}\ \bibnamefont {Dewhurst}}, \bibinfo {author}
  {\bibfnamefont {Y.}~\bibnamefont {Onose}}, \bibinfo {author} {\bibfnamefont
  {Y.}~\bibnamefont {Endoh}}, \bibinfo {author} {\bibfnamefont
  {Y.}~\bibnamefont {Tokura}}, \ and\ \bibinfo {author} {\bibfnamefont
  {C.}~\bibnamefont {Pappas}},\ }\href@noop {} {\bibfield  {journal} {\bibinfo
  {journal} {Physical Review B}\ }\textbf {\bibinfo {volume} {94}},\ \bibinfo
  {pages} {104406} (\bibinfo {year} {2016})}\BibitemShut {NoStop}%
\bibitem [{\citenamefont {Milde}\ \emph {et~al.}(2013)\citenamefont {Milde},
  \citenamefont {K{\"o}hler}, \citenamefont {Seidel}, \citenamefont {Eng},
  \citenamefont {Bauer}, \citenamefont {Chacon}, \citenamefont {Kindervater},
  \citenamefont {M{\"u}hlbauer}, \citenamefont {Pfleiderer}, \citenamefont
  {Buhrandt} \emph {et~al.}}]{milde2013}%
  \BibitemOpen
  \bibfield  {author} {\bibinfo {author} {\bibfnamefont {P.}~\bibnamefont
  {Milde}}, \bibinfo {author} {\bibfnamefont {D.}~\bibnamefont {K{\"o}hler}},
  \bibinfo {author} {\bibfnamefont {J.}~\bibnamefont {Seidel}}, \bibinfo
  {author} {\bibfnamefont {L.}~\bibnamefont {Eng}}, \bibinfo {author}
  {\bibfnamefont {A.}~\bibnamefont {Bauer}}, \bibinfo {author} {\bibfnamefont
  {A.}~\bibnamefont {Chacon}}, \bibinfo {author} {\bibfnamefont
  {J.}~\bibnamefont {Kindervater}}, \bibinfo {author} {\bibfnamefont
  {S.}~\bibnamefont {M{\"u}hlbauer}}, \bibinfo {author} {\bibfnamefont
  {C.}~\bibnamefont {Pfleiderer}}, \bibinfo {author} {\bibfnamefont
  {S.}~\bibnamefont {Buhrandt}},  \emph {et~al.},\ }\href@noop {} {\bibfield
  {journal} {\bibinfo  {journal} {Science}\ }\textbf {\bibinfo {volume}
  {340}},\ \bibinfo {pages} {1076} (\bibinfo {year} {2013})}\BibitemShut
  {NoStop}%
\bibitem [{\citenamefont {Watanabe}\ \emph {et~al.}(1985)\citenamefont
  {Watanabe}, \citenamefont {Tazuke},\ and\ \citenamefont
  {Nakajima}}]{watanabe1985}%
  \BibitemOpen
  \bibfield  {author} {\bibinfo {author} {\bibfnamefont {H.}~\bibnamefont
  {Watanabe}}, \bibinfo {author} {\bibfnamefont {i.}~\bibnamefont {Tazuke}}, \
  and\ \bibinfo {author} {\bibfnamefont {H.}~\bibnamefont {Nakajima}},\
  }\href@noop {} {\bibfield  {journal} {\bibinfo  {journal} {Journal of the
  Physical Society of Japan}\ }\textbf {\bibinfo {volume} {54}},\ \bibinfo
  {pages} {3978} (\bibinfo {year} {1985})}\BibitemShut {NoStop}%
\bibitem [{\citenamefont {Chattopadhyay}\ \emph {et~al.}(2002)\citenamefont
  {Chattopadhyay}, \citenamefont {Roy},\ and\ \citenamefont
  {Chaudhary}}]{chattopadhyay2002}%
  \BibitemOpen
  \bibfield  {author} {\bibinfo {author} {\bibfnamefont {M.~K.}\ \bibnamefont
  {Chattopadhyay}}, \bibinfo {author} {\bibfnamefont {S.~B.}\ \bibnamefont
  {Roy}}, \ and\ \bibinfo {author} {\bibfnamefont {S.}~\bibnamefont
  {Chaudhary}},\ }\href@noop {} {\bibfield  {journal} {\bibinfo  {journal}
  {Physical Review B}\ }\textbf {\bibinfo {volume} {65}},\ \bibinfo {pages}
  {132409} (\bibinfo {year} {2002})}\BibitemShut {NoStop}%
\bibitem [{\citenamefont {Grigoriev}\ \emph {et~al.}(2007)\citenamefont
  {Grigoriev}, \citenamefont {Dyadkin}, \citenamefont {Menzel}, \citenamefont
  {Schoenes}, \citenamefont {Chetverikov}, \citenamefont {Okorokov},
  \citenamefont {Eckerlebe},\ and\ \citenamefont {Maleyev}}]{grigoriev2007}%
  \BibitemOpen
  \bibfield  {author} {\bibinfo {author} {\bibfnamefont {S.~V.}\ \bibnamefont
  {Grigoriev}}, \bibinfo {author} {\bibfnamefont {V.~A.}\ \bibnamefont
  {Dyadkin}}, \bibinfo {author} {\bibfnamefont {D.}~\bibnamefont {Menzel}},
  \bibinfo {author} {\bibfnamefont {J.}~\bibnamefont {Schoenes}}, \bibinfo
  {author} {\bibfnamefont {Y.~O.}\ \bibnamefont {Chetverikov}}, \bibinfo
  {author} {\bibfnamefont {A.~I.}\ \bibnamefont {Okorokov}}, \bibinfo {author}
  {\bibfnamefont {H.}~\bibnamefont {Eckerlebe}}, \ and\ \bibinfo {author}
  {\bibfnamefont {S.~V.}\ \bibnamefont {Maleyev}},\ }\href@noop {} {\bibfield
  {journal} {\bibinfo  {journal} {Physical Review B}\ }\textbf {\bibinfo
  {volume} {76}},\ \bibinfo {pages} {224424} (\bibinfo {year}
  {2007})}\BibitemShut {NoStop}%
\bibitem [{\citenamefont {Ou-Yang}\ \emph {et~al.}(2015)\citenamefont
  {Ou-Yang}, \citenamefont {Shu}, \citenamefont {Hu},\ and\ \citenamefont
  {Chou}}]{ou-yang2015}%
  \BibitemOpen
  \bibfield  {author} {\bibinfo {author} {\bibfnamefont {T.~Y.}\ \bibnamefont
  {Ou-Yang}}, \bibinfo {author} {\bibfnamefont {G.~J.}\ \bibnamefont {Shu}},
  \bibinfo {author} {\bibfnamefont {C.~D.}\ \bibnamefont {Hu}}, \ and\ \bibinfo
  {author} {\bibfnamefont {F.~C.}\ \bibnamefont {Chou}},\ }\href@noop {}
  {\bibfield  {journal} {\bibinfo  {journal} {Journal of Applied Physics}\
  }\textbf {\bibinfo {volume} {117}},\ \bibinfo {pages} {123903} (\bibinfo
  {year} {2015})}\BibitemShut {NoStop}%
\bibitem [{\citenamefont {Takeda}\ \emph {et~al.}(2009)\citenamefont {Takeda},
  \citenamefont {Endoh}, \citenamefont {Kakurai}, \citenamefont {Onose},
  \citenamefont {Suzuki},\ and\ \citenamefont {Tokura}}]{takeda2009}%
  \BibitemOpen
  \bibfield  {author} {\bibinfo {author} {\bibfnamefont {M.}~\bibnamefont
  {Takeda}}, \bibinfo {author} {\bibfnamefont {Y.}~\bibnamefont {Endoh}},
  \bibinfo {author} {\bibfnamefont {K.}~\bibnamefont {Kakurai}}, \bibinfo
  {author} {\bibfnamefont {Y.}~\bibnamefont {Onose}}, \bibinfo {author}
  {\bibfnamefont {J.}~\bibnamefont {Suzuki}}, \ and\ \bibinfo {author}
  {\bibfnamefont {Y.}~\bibnamefont {Tokura}},\ }\href@noop {} {\bibfield
  {journal} {\bibinfo  {journal} {Journal of the Physical Society of Japan}\
  }\textbf {\bibinfo {volume} {78}},\ \bibinfo {pages} {093704} (\bibinfo
  {year} {2009})}\BibitemShut {NoStop}%
\bibitem [{\citenamefont {Bauer}\ and\ \citenamefont
  {Pfleiderer}(2012)}]{bauer2012}%
  \BibitemOpen
  \bibfield  {author} {\bibinfo {author} {\bibfnamefont {A.}~\bibnamefont
  {Bauer}}\ and\ \bibinfo {author} {\bibfnamefont {C.}~\bibnamefont
  {Pfleiderer}},\ }\href@noop {} {\bibfield  {journal} {\bibinfo  {journal}
  {Physical Review B}\ }\textbf {\bibinfo {volume} {85}},\ \bibinfo {pages}
  {214418} (\bibinfo {year} {2012})}\BibitemShut {NoStop}%
\bibitem [{\citenamefont {Qian}\ \emph {et~al.}(2016)\citenamefont {Qian},
  \citenamefont {Wilhelm}, \citenamefont {Aqeel}, \citenamefont {Palstra},
  \citenamefont {Lefering}, \citenamefont {Bruck},\ and\ \citenamefont
  {Pappas}}]{qian2016}%
  \BibitemOpen
  \bibfield  {author} {\bibinfo {author} {\bibfnamefont {F.}~\bibnamefont
  {Qian}}, \bibinfo {author} {\bibfnamefont {H.}~\bibnamefont {Wilhelm}},
  \bibinfo {author} {\bibfnamefont {A.}~\bibnamefont {Aqeel}}, \bibinfo
  {author} {\bibfnamefont {T.~T.~M.}\ \bibnamefont {Palstra}}, \bibinfo
  {author} {\bibfnamefont {A.~J.~E.}\ \bibnamefont {Lefering}}, \bibinfo
  {author} {\bibfnamefont {E.~H.}\ \bibnamefont {Bruck}}, \ and\ \bibinfo
  {author} {\bibfnamefont {C.}~\bibnamefont {Pappas}},\ }\href@noop {}
  {\bibfield  {journal} {\bibinfo  {journal} {Physical Review B}\ }\textbf
  {\bibinfo {volume} {94}},\ \bibinfo {pages} {064418} (\bibinfo {year}
  {2016})}\BibitemShut {NoStop}%
\bibitem [{\citenamefont {Thessieu}\ \emph {et~al.}(1997)\citenamefont
  {Thessieu}, \citenamefont {Pfleiderer}, \citenamefont {Stepanov},\ and\
  \citenamefont {Flouquet}}]{thessieu1997}%
  \BibitemOpen
  \bibfield  {author} {\bibinfo {author} {\bibfnamefont {C.}~\bibnamefont
  {Thessieu}}, \bibinfo {author} {\bibfnamefont {C.}~\bibnamefont
  {Pfleiderer}}, \bibinfo {author} {\bibfnamefont {A.}~\bibnamefont
  {Stepanov}}, \ and\ \bibinfo {author} {\bibfnamefont {J.}~\bibnamefont
  {Flouquet}},\ }\href@noop {} {\bibfield  {journal} {\bibinfo  {journal}
  {Journal of Physics: Condensed Matter}\ }\textbf {\bibinfo {volume} {9}},\
  \bibinfo {pages} {6677} (\bibinfo {year} {1997})}\BibitemShut {NoStop}%
\bibitem [{\citenamefont {Wilhelm}\ \emph {et~al.}(2011)\citenamefont
  {Wilhelm}, \citenamefont {Baenitz}, \citenamefont {Schmidt}, \citenamefont
  {R{\"o}{\ss}ler}, \citenamefont {Leonov},\ and\ \citenamefont
  {Bogdanov}}]{wilhelm2011}%
  \BibitemOpen
  \bibfield  {author} {\bibinfo {author} {\bibfnamefont {H.}~\bibnamefont
  {Wilhelm}}, \bibinfo {author} {\bibfnamefont {M.}~\bibnamefont {Baenitz}},
  \bibinfo {author} {\bibfnamefont {M.}~\bibnamefont {Schmidt}}, \bibinfo
  {author} {\bibfnamefont {U.~K.}\ \bibnamefont {R{\"o}{\ss}ler}}, \bibinfo
  {author} {\bibfnamefont {A.~A.}\ \bibnamefont {Leonov}}, \ and\ \bibinfo
  {author} {\bibfnamefont {A.~N.}\ \bibnamefont {Bogdanov}},\ }\href@noop {}
  {\bibfield  {journal} {\bibinfo  {journal} {Physical Review Letters}\
  }\textbf {\bibinfo {volume} {107}},\ \bibinfo {pages} {127203} (\bibinfo
  {year} {2011})}\BibitemShut {NoStop}%
\bibitem [{\citenamefont {{\v{Z}}ivkovi{\'c}}\ \emph
  {et~al.}(2014)\citenamefont {{\v{Z}}ivkovi{\'c}}, \citenamefont {White},
  \citenamefont {R{\o}nnow}, \citenamefont {Pr{\v{s}}a},\ and\ \citenamefont
  {Berger}}]{zivkovic2014}%
  \BibitemOpen
  \bibfield  {author} {\bibinfo {author} {\bibfnamefont {I.}~\bibnamefont
  {{\v{Z}}ivkovi{\'c}}}, \bibinfo {author} {\bibfnamefont {J.~S.}\ \bibnamefont
  {White}}, \bibinfo {author} {\bibfnamefont {H.~M.}\ \bibnamefont
  {R{\o}nnow}}, \bibinfo {author} {\bibfnamefont {K.}~\bibnamefont
  {Pr{\v{s}}a}}, \ and\ \bibinfo {author} {\bibfnamefont {H.}~\bibnamefont
  {Berger}},\ }\href@noop {} {\bibfield  {journal} {\bibinfo  {journal}
  {Physical Review B}\ }\textbf {\bibinfo {volume} {89}},\ \bibinfo {pages}
  {060401} (\bibinfo {year} {2014})}\BibitemShut {NoStop}%
\bibitem [{\citenamefont {Bauer}\ \emph {et~al.}(2010)\citenamefont {Bauer},
  \citenamefont {Neubauer}, \citenamefont {Franz}, \citenamefont {M{\"u}nzer},
  \citenamefont {Garst},\ and\ \citenamefont {Pfleiderer}}]{bauer2010}%
  \BibitemOpen
  \bibfield  {author} {\bibinfo {author} {\bibfnamefont {A.}~\bibnamefont
  {Bauer}}, \bibinfo {author} {\bibfnamefont {A.}~\bibnamefont {Neubauer}},
  \bibinfo {author} {\bibfnamefont {C.}~\bibnamefont {Franz}}, \bibinfo
  {author} {\bibfnamefont {W.}~\bibnamefont {M{\"u}nzer}}, \bibinfo {author}
  {\bibfnamefont {M.}~\bibnamefont {Garst}}, \ and\ \bibinfo {author}
  {\bibfnamefont {C.}~\bibnamefont {Pfleiderer}},\ }\href@noop {} {\bibfield
  {journal} {\bibinfo  {journal} {Physical Review B}\ }\textbf {\bibinfo
  {volume} {82}},\ \bibinfo {pages} {064404} (\bibinfo {year}
  {2010})}\BibitemShut {NoStop}%
\bibitem [{\citenamefont {Levati{\'c}}\ \emph {et~al.}(2016)\citenamefont
  {Levati{\'c}}, \citenamefont {Pop{\v{c}}evi{\'c}}, \citenamefont
  {{\v{S}}urija}, \citenamefont {Kruchkov}, \citenamefont {Berger},
  \citenamefont {Magrez}, \citenamefont {White}, \citenamefont {R{\o}nnow},\
  and\ \citenamefont {{\v{Z}}ivkovi{\'c}}}]{levatic2016}%
  \BibitemOpen
  \bibfield  {author} {\bibinfo {author} {\bibfnamefont {I.}~\bibnamefont
  {Levati{\'c}}}, \bibinfo {author} {\bibfnamefont {P.}~\bibnamefont
  {Pop{\v{c}}evi{\'c}}}, \bibinfo {author} {\bibfnamefont {V.}~\bibnamefont
  {{\v{S}}urija}}, \bibinfo {author} {\bibfnamefont {A.}~\bibnamefont
  {Kruchkov}}, \bibinfo {author} {\bibfnamefont {H.}~\bibnamefont {Berger}},
  \bibinfo {author} {\bibfnamefont {A.}~\bibnamefont {Magrez}}, \bibinfo
  {author} {\bibfnamefont {J.}~\bibnamefont {White}}, \bibinfo {author}
  {\bibfnamefont {H.}~\bibnamefont {R{\o}nnow}}, \ and\ \bibinfo {author}
  {\bibfnamefont {I.}~\bibnamefont {{\v{Z}}ivkovi{\'c}}},\ }\href@noop {}
  {\bibfield  {journal} {\bibinfo  {journal} {Scientific reports}\ }\textbf
  {\bibinfo {volume} {6}},\ \bibinfo {pages} {21347} (\bibinfo {year}
  {2016})}\BibitemShut {NoStop}%
\bibitem [{\citenamefont {Campbell}(1986)}]{campbell1986}%
  \BibitemOpen
  \bibfield  {author} {\bibinfo {author} {\bibfnamefont {I.~A.}\ \bibnamefont
  {Campbell}},\ }\href@noop {} {\bibfield  {journal} {\bibinfo  {journal}
  {Physical Review B}\ }\textbf {\bibinfo {volume} {33}},\ \bibinfo {pages}
  {3587} (\bibinfo {year} {1986})}\BibitemShut {NoStop}%
\bibitem [{\citenamefont {Huser}\ \emph {et~al.}(1986)\citenamefont {Huser},
  \citenamefont {Van~Duyneveldt}, \citenamefont {Nieuwenhuys},\ and\
  \citenamefont {Mydosh}}]{huser1986}%
  \BibitemOpen
  \bibfield  {author} {\bibinfo {author} {\bibfnamefont {D.}~\bibnamefont
  {Huser}}, \bibinfo {author} {\bibfnamefont {A.~J.}\ \bibnamefont
  {Van~Duyneveldt}}, \bibinfo {author} {\bibfnamefont {G.~J.}\ \bibnamefont
  {Nieuwenhuys}}, \ and\ \bibinfo {author} {\bibfnamefont {J.~A.}\ \bibnamefont
  {Mydosh}},\ }\href@noop {} {\bibfield  {journal} {\bibinfo  {journal}
  {Journal of Physics C: Solid State Physics}\ }\textbf {\bibinfo {volume}
  {19}},\ \bibinfo {pages} {3697} (\bibinfo {year} {1986})}\BibitemShut
  {NoStop}%
\bibitem [{\citenamefont {Dekker}\ \emph {et~al.}(1989)\citenamefont {Dekker},
  \citenamefont {Arts}, \citenamefont {de~Wijn}, \citenamefont {van
  Duyneveldt},\ and\ \citenamefont {Mydosh}}]{dekker1989}%
  \BibitemOpen
  \bibfield  {author} {\bibinfo {author} {\bibfnamefont {C.}~\bibnamefont
  {Dekker}}, \bibinfo {author} {\bibfnamefont {A.~F.~M.}\ \bibnamefont {Arts}},
  \bibinfo {author} {\bibfnamefont {H.~W.}\ \bibnamefont {de~Wijn}}, \bibinfo
  {author} {\bibfnamefont {A.~J.}\ \bibnamefont {van Duyneveldt}}, \ and\
  \bibinfo {author} {\bibfnamefont {J.~A.}\ \bibnamefont {Mydosh}},\
  }\href@noop {} {\bibfield  {journal} {\bibinfo  {journal} {Physical Review
  B}\ }\textbf {\bibinfo {volume} {40}},\ \bibinfo {pages} {11243} (\bibinfo
  {year} {1989})}\BibitemShut {NoStop}%
\bibitem [{\citenamefont {Qian}(2016)}]{qian2016MnSi}%
  \BibitemOpen
  \bibfield  {author} {\bibinfo {author} {\bibfnamefont {F.}~\bibnamefont
  {Qian}},\ }\href@noop {} {\bibfield  {journal} {\bibinfo  {journal} {Private
  Communication}\ } (\bibinfo {year} {2016})}\BibitemShut {NoStop}%
\bibitem [{Note1()}]{Note1}%
  \BibitemOpen
  \bibinfo {note} {At a frequency of 0.1~Hz, the SQUID has difficulties
  stabilizing the phase in a region from $T$ = 42 to 50~K and fields below
  50~mT, which is likely caused by the sample.}\BibitemShut {Stop}%
\bibitem [{\citenamefont {Tsuruta}\ \emph {et~al.}(2016)\citenamefont
  {Tsuruta}, \citenamefont {Mito}, \citenamefont {Kousaka}, \citenamefont
  {Akimutsu}, \citenamefont {Kishine},\ and\ \citenamefont
  {Inoue}}]{tsuruta2016}%
  \BibitemOpen
  \bibfield  {author} {\bibinfo {author} {\bibfnamefont {K.}~\bibnamefont
  {Tsuruta}}, \bibinfo {author} {\bibfnamefont {M.}~\bibnamefont {Mito}},
  \bibinfo {author} {\bibfnamefont {Y.}~\bibnamefont {Kousaka}}, \bibinfo
  {author} {\bibfnamefont {J.}~\bibnamefont {Akimutsu}}, \bibinfo {author}
  {\bibfnamefont {J.}~\bibnamefont {Kishine}}, \ and\ \bibinfo {author}
  {\bibfnamefont {K.}~\bibnamefont {Inoue}},\ }\href@noop {} {\bibfield
  {journal} {\bibinfo  {journal} {Private Communication}\ } (\bibinfo {year}
  {2016})}\BibitemShut {NoStop}%
\bibitem [{Note2()}]{Note2}%
  \BibitemOpen
  \bibinfo {note} {The cooling rate reported here for FC refers to the average
  cooling rate between $T$ = 40~K and 30~K.}\BibitemShut {Stop}%
\end{thebibliography}%

\end{document}